\documentclass{aa}

\usepackage{natbib}
\usepackage{subfig}
\bibpunct{(}{)}{;}{a}{}{,} 
\usepackage{upgreek}
\usepackage{graphicx}
\usepackage{txfonts}
\usepackage{threeparttablex} 
\usepackage{lscape} 
\usepackage{url}
\usepackage{tablefootnote}

\usepackage{xcolor}
\usepackage[normalem]{ulem}

\def\deg  {\ifmmode {^\circ}\else {$^\circ$}\fi}

\definecolor{malachite}{rgb}{0.34, 0.7, 0.22}

\def\ag{$A _\mathrm{G}$}

\def\kms     {km~s$^{-1}$}

\newcommand{\irasum}{IRAS  100-$\mathrm{\upmu}$m}

\newcommand{\planck}{Planck  857-GHz}

\begin{document}
 
   \title{The Distances to Molecular Clouds at High Galactic Latitudes based on \textit{Gaia} DR2}
 
   \author{Qing-Zeng Yan\inst{1,2}, Bo Zhang\inst{2}, Ye Xu\inst{1}, Sufen Guo\inst{2,3}, J. P. Macquart\inst{4}, Zheng-Hong Tang\inst{2,3},    A. J. Walsh\inst{5}}

   \institute{    Purple Mountain Observatory, Nanjing 210008, China
     \email{qzyan@pmo.ac.cn}
     \and
       Shanghai Astronomical Observatory, Chinese Academy of Sciences, Shanghai 200030, China
            \email{zb@shao.ac.cn}
      \and
  School of Astronomy and Space Science, University of Chinese Academy of Sciences, 19A Yuquanlu, Beijing 100049, China
        \and
      International Centre for Radio Astronomy Research, Curtin University, GPO Box U1987, Perth WA 6845, Australia
       \and
   Department of Physics and Astronomy and MQ Research Centre in Astronomy, Astrophysics and Astrophotonics, Macquarie University, NSW 2109, Australia}

   \date{Accepted }
   
\titlerunning{Molecular cloud distances}
\authorrunning{Yan et al}

  \abstract
   {  We report the distances of   molecular clouds at high Galactic latitudes ($|b|>10$\deg) derived from parallax and G band extinction  (\ag) measurements  in the second \textit{Gaia} data release, \textit{Gaia} DR2.    Aided by   Bayesian analyses, we determined distances by identifying the  breakpoint in the extinction \ag\ towards molecular clouds  and using the extinction \ag\ of  \textit{Gaia} stars around molecular clouds to confirm the breakpoint.   We use nearby star-forming regions, such as Orion, Taurus, Cepheus, and Perseus,  whose distances are well-known to examine the reliability of our method. By comparing with previous results, we found  that the molecular cloud distances derived from this method are reliable. The systematic error in the distances is approximately 5\%.  In total, 52  molecular clouds have their distances well determined, most of which  are at high Galactic latitudes,  and we provide  reliable distances for 13 molecular clouds for the first time. } 
\keywords{dust, extinction - ISM: clouds}

   \maketitle
%


\section{Introduction} \label{sec:intro}

Determining the distances of molecular clouds, the birthplaces  of stars, is usually difficult.  Many distance measurement methods, such as the photometric  parallax method and the period-luminosity relation, are not applicable to molecular clouds.  However, the distance relates directly to  many important properties  of molecular clouds, especially the mass and  size, making their physical states and relationships with Galactic spiral arms largely uncertain. 
 
There are a few approaches to derive   molecular cloud distances, but they usually either yield large uncertainties or are not applicable to diffuse and translucent molecular clouds at high Galactic latitudes. For instance,  maser astrometry with multiple-epoch Very Long Baseline  Interferometry (VLBI) measurements \citep{2009ApJ...693..413X, 2013ApJ...775...79Z, 2017ApJ...834..143O}, cannot be performed towards  molecular clouds that lack masers. Furthermore,    kinematic distance estimates based on modeled rotation curves \citep{2009ApJ...699.1153R, 2016ApJ...822...52R} suffer from large uncertainties and ambiguities.

Optical extinction provides another approach to determine the cloud distances.  However, the optical extinction derived from counting stars towards molecular clouds ~\citep{1937dss..book.....B,1986A&A...168..271M} involves large  uncertainties. A more straightforward but sophisticated way is to examine the variation of optical extinction with respect to distance along the line of sight \citep[see][]{1998A&A...338..897K}. For example, using this method, \citet{2014ApJ...786...29S} provides a large distance catalog of 18 well-known star-forming regions and  108 high-Galactic-latitude molecular clouds cataloged by \citet[hereafter denoted as  MBM]{1985ApJ...295..402M}. For each star, they derived its distance and extinction   simultaneously  from Pan-STARRS1 \citep{2010SPIE.7733E..0EK,2014ApJ...783..114G} photometry  and subsequently estimated the molecular cloud distances  according to the breakpoint of the extinction. However, as \citet{2014ApJ...786...29S}  have pointed out, their distances may have a $\sim$10\% systematic uncertainty. 
 

The release of  the second \textit{Gaia} data~\citep{2016A&A...595A...1G, 2018A&A...616A...1G}, \textit{Gaia} DR2,    providing  parallaxes for about  1.3 billion stars,  a large proportion of which have G band extinction (\ag) measurements,  has advanced this approach substantially. The correspondence between the   extinction \ag\ map  \citep{2018A&A...616A..17A} and the  distribution of molecular clouds \citep{2001ApJ...547..792D}  suggests that the extinction \ag\ is  capable of detecting molecular  clouds.  Although, as warned by \citet{2018A&A...616A...1G},  the extinction \ag\ has large uncertainties, it is found to be reliable on average \citep[see][]{2018A&A...616A...8A}.

  \textit{Gaia} DR2 permits the analysis of the extinction caused by molecular clouds more precisely than previously possible and offers an independent means of   examining  distance estimates of \citet{2014ApJ...786...29S}.   The rest of this paper is organized as follows. In \S2 and \S3, we introduce the methodology of reducing \textit{Gaia} data and the process of deriving molecular cloud distances. We present the distance  catalog  in \S4 and compare the results with previous studies in \S 5.  The conclusions are  summarized  in   \S6. 
  
\section{  The  \textit{Gaia} DR2 and \planck\  Data} \label{sec:data}
We primarily investigate diffuse molecular clouds at  $|b|>10$\deg, which are less likely contaminated by other molecular clouds either adjacent or  overlapping along the line of sight. Despite   being located at high Galactic latitudes, they are usually less than 1 kpc from the Sun and are still located  in the Galactic plane \citep{1996ApJS..106..447M,2014ApJ...786...29S}, They tend  to  occupy  large areas on the sky, making them able to optically extinct a large number of stars. Consequently, the extinction \ag\ imposed  on stars behind  molecular clouds  is able to be investigated statistically. 

We selected \textit{Gaia} DR2 stars according to their parallaxes and extinction \ag.  First, we removed those stars with parallaxes $<$ 0.5 mas, corresponding to an upper distance threshold of 2 kpc. Molecular clouds at high Galactic latitudes are usually near the Sun ($<$ 1 kpc), and a 2-kpc cutoff suffices for our study. If the relative errors of stellar parallaxes exceed 20\%, the corresponding distances would differ from the reciprocal of their  parallaxes by a large amount \citep{2015PASP..127..994B}, and consequently, we only kept those stars whose relative parallax errors are less than 20\%.  As to the extinction \ag, we require \ag\  $>$ 0, which rejects 13 stars. In total, we have 30,259,242 \textit{Gaia} stars meet the criteria. 

The  extinction \ag\ error of a single star, $\Delta\mathrm{A}_G$, is estimated with
\begin{equation}
\Delta A_\mathrm{G} = \rm \frac{1}{2}\left( 84th\_percentile -16th\_percentile\right).
\label{equ:agstd}
\end{equation}
where $\rm 84th\_percentile$ and $\rm 16th\_percentile$ are provided in  \textit{Gaia} DR2. To derive the distance and its standard deviation, we drew  10000 samples from the  Gaussian distribution $\mathcal{N}(\varpi, \Delta \varpi )$, where $\varpi$ and $\Delta \varpi$ are the parallax and  its error in \textit{Gaia} DR2, and the  mean and standard deviation of the distance correspond to the mean and standard deviation of the reciprocal of the 10000 samples, respectively.

We used \planck\ images  \citep{2014A&A...571A...1P} to trace the molecular clouds rather than \irasum\ and CO spectral lines.  Although the \irasum\ images \citep{1998ApJ...500..525S, 2005ApJS..157..302M} have a comparable spatial resolution ($5$\arcmin) with  the \planck\ survey, the sensitivity of  the  \planck\ survey is higher, and consequently, the \planck\ survey classifies \textit{Gaia} stars more accurately and produces slightly better results, details of which  are discussed in \S\ref{sec:planckvsiras}.  \planck\  data  is  more complete than  CO survey data~\citep{2001ApJ...547..792D}  at high Galactic latitudes, and  although  \planck\ emission cannot distinguish   molecular cloud components that overlap each other along the line of sight,  this situation is not severe at high Galactic latitudes, and usually only one nearby  molecular cloud component is present towards one direction. The \planck\ images are only used to classify \textit{Gaia} DR2 stars, not to build dust models.

 \section{METHOD} \label{sec:method}
 
In this section we describe the method of deriving molecular cloud distances  with the \textit{Gaia} DR2   and \planck\ data. Principally, molecular clouds increase the extinction \ag\ of all stars behind them, thus producing breakpoints in the extinction \ag\ along the line of sight. Consequently,  identifying the breakpoints in the extinction \ag\ is the essential point in our method.

\begin{figure}[ht!]
   \centering
 \includegraphics{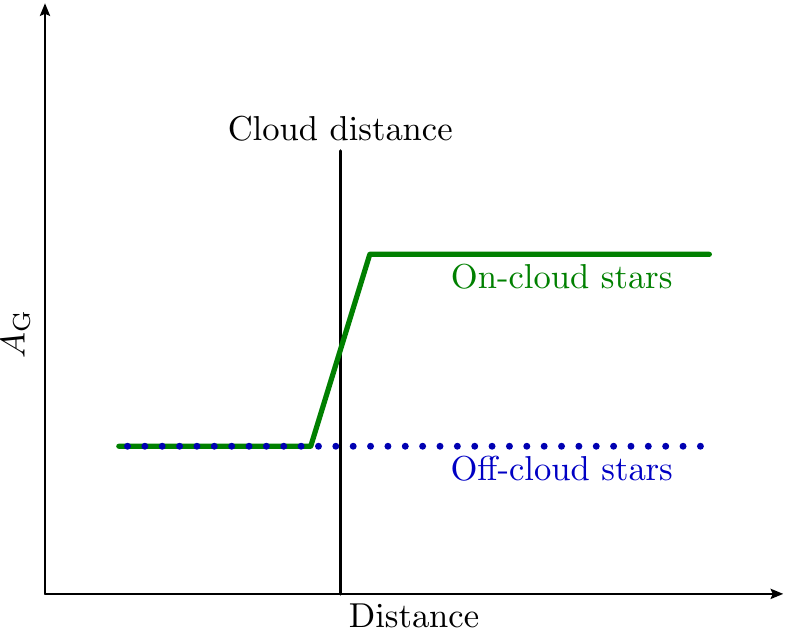} 
 \caption{ The  extinction \ag\ of on- (green) and off-cloud (blue) stars. The  vertical black line marks the distance of the molecular cloud. \label{fig:principle}}
 \end{figure}

First, we collect  two classes of \textit{Gaia} stars, on- and off-cloud stars,  based on \planck\ emission.  On-cloud stars,  showing breakpoints  in the extinction \ag,  are those \textit{Gaia} stars towards molecular clouds.  We define as ``off-cloud stars'' those \textit{Gaia} stars around molecular clouds, and because  they are not  affected by molecular clouds, off-cloud stars have no breakpoints in the extinction \ag. Practically, the breakpoints are determined with only on-cloud stars, while off-cloud stars are used to confirm the breakpoints by eye.  The schematic diagram of Figure \ref{fig:principle} depicts the extinction \ag\ feature of on- (green) and off-cloud (blue) stars.


In the second  step, the distances are determined using Bayesian inference with on-cloud stars and are subsequently confirmed with off-cloud stars by eye. We use two molecular clouds, Taurus \citep{2010ApJ...721..686P, 2016MNRAS.463.1008W} and Gemini \citep{2015AJ....150...60L}, to illustrate the process of determining  distances.





  \begin{figure*}[ht]
  \centering
 \subfloat[]{\includegraphics[width=0.48\textwidth]{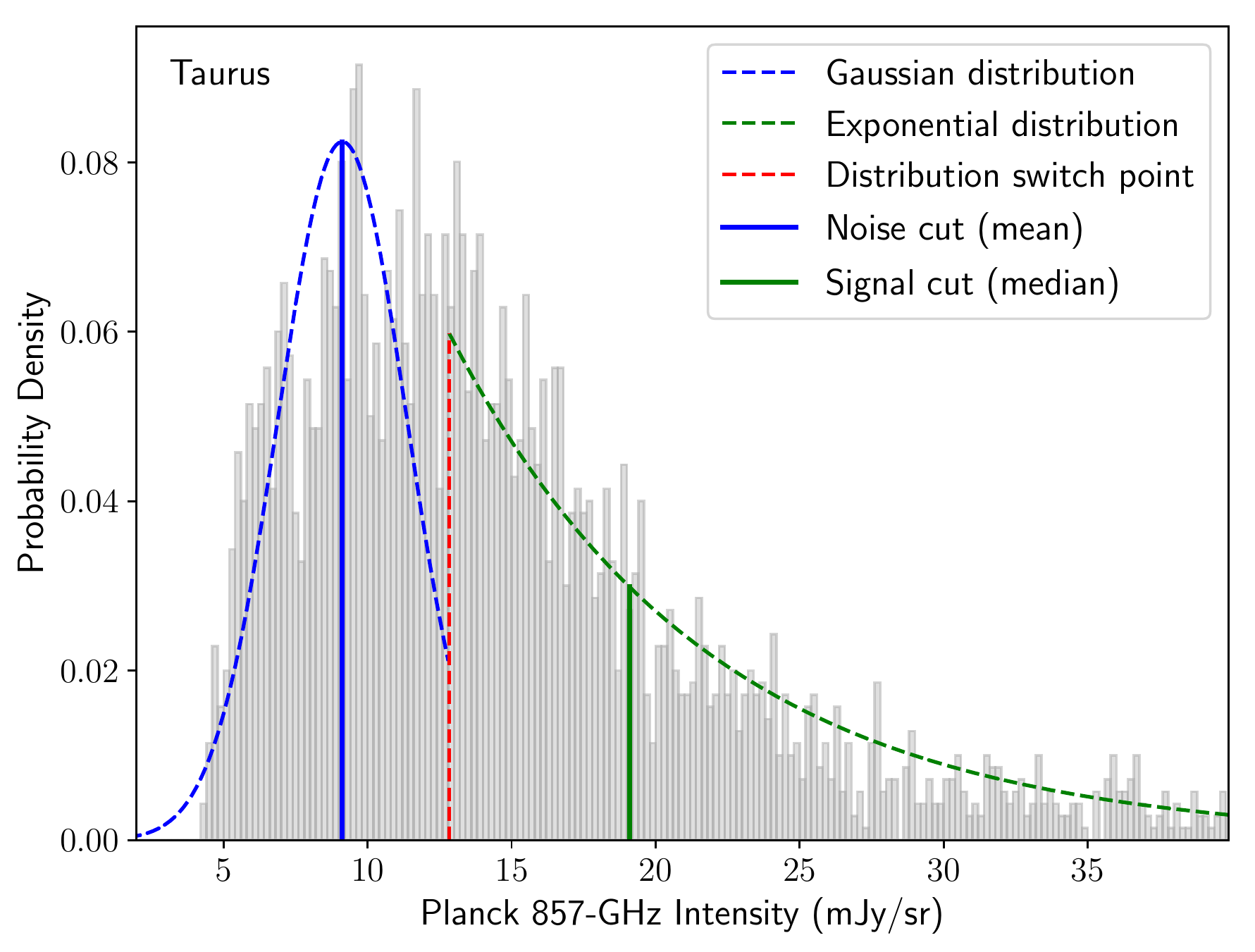}}
  \subfloat[]{\includegraphics[width=0.48\textwidth]{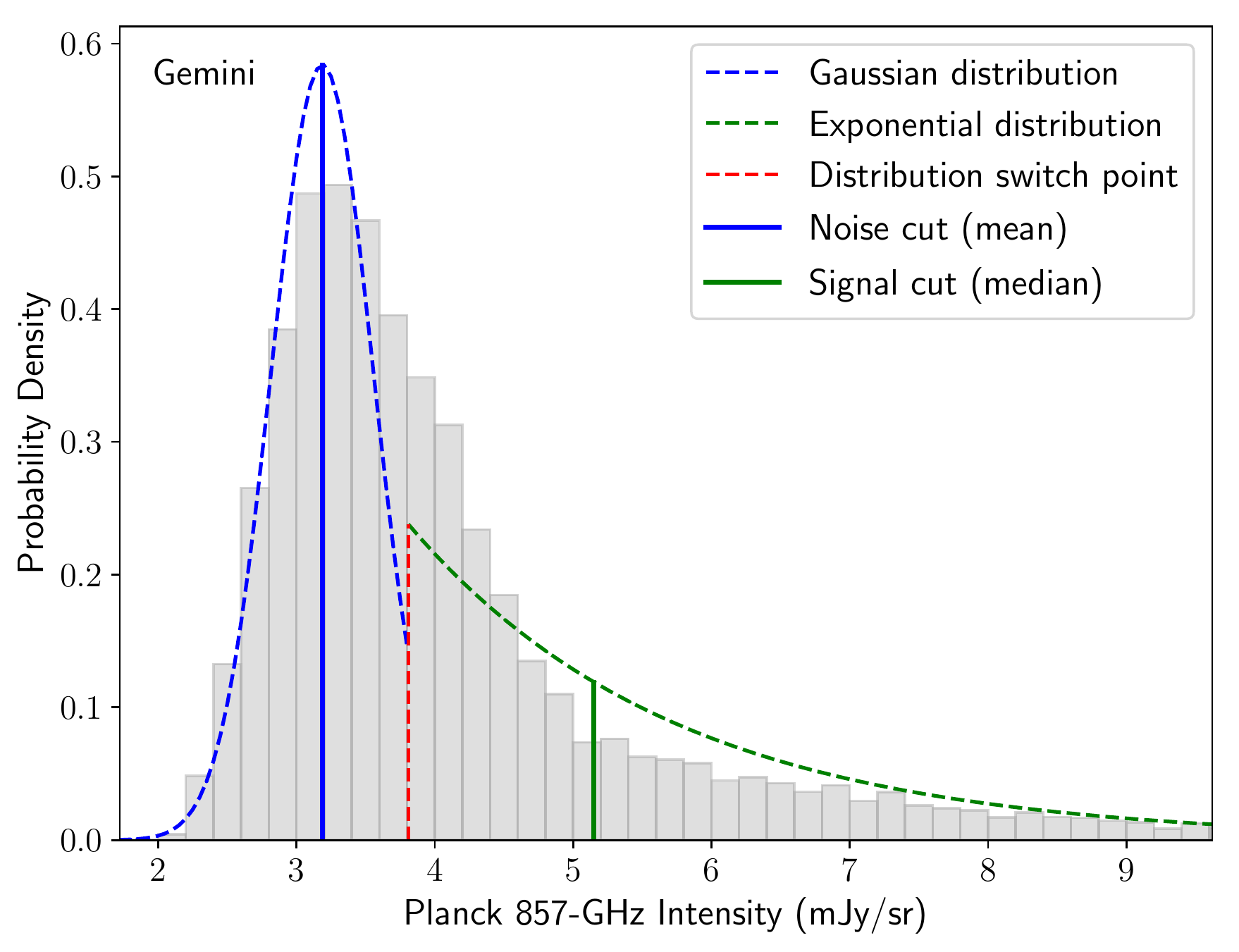}}
\caption{ Determining the noise and signal levels to classify on- and off-cloud  stars for the Taurus (a) and  Gemini (b) molecular clouds. The blue and green dashed lines represent the Gaussian and exponential distribution, respectively, and the dashed vertical red lines mark the switch point between the two distributions. We use the Gaussian mean (the solid vertical blue lines) for  the noise cutoff level and  the exponential median (the solid vertical green lines)  for the signal cutoff level.  \label{fig:snlevel}}
\end{figure*}


\subsection{On- and off-cloud stars}\label{sub:snlevel}
 
The on- and off-cloud stars are classified according to \planck\ intensity thresholds.  The on-cloud stars are those whose \planck\ emission is stronger than an intensity threshold  (the signal  level),  while  towards off-cloud stars, the \planck\ emission is  fainter than a  lower threshold (the noise  level).  We determined the two thresholds  through  fitting  a mixed  distribution combining two functions,   Gaussian + exponential, as described below. 


First, we manually drew a box region for each molecular cloud, and only those \textit{Gaia} stars in this box region were considered. This region contains at least part of the molecular cloud and incorporates an extra nearby region. This extra  region, where the  \planck\ emission is significantly lower than  the  molecular cloud region, contains off-cloud stars. Noteworthy is that these box regions are not necessarily the same with traditional boundaries of molecular clouds, i.e., they may be larger or smaller than the entire molecular clouds as long as those regions contain sufficient \textit{Gaia} stars.


Secondly, we assigned each \textit{Gaia} star a \planck\ emission value according to the \planck\ data.  As the histograms  demonstrate  in Figure~\ref{fig:snlevel},  the \planck\ emission roughly contains two components: (1) a background noise  that has approximately Gaussian distribution in intensity; (2) the  molecular cloud emission that resembles an exponential distribution. 

 We use four parameters to model this mixed distribution: the mean ($\upmu$) and the  standard deviation  ($\upsigma$) of the Gaussian part, a switch point ($\rm SP$), and the rate ($\uplambda$) of the exponential distribution. The cumulative distribution function (CDF) of the mixed distribution (the likelihood) is 
\begin{equation}
  \int_{-\infty}^{x} p\left(I  |\upmu,\upsigma,\rm SP,\uplambda\right) \mathrm{d} I=  \left\{
             \begin{array}{l}
           \int_{-\infty}^{x} \frac{1}{\sqrt{2\uppi} \upsigma} \exp\left({-\frac { (I- \upmu)^2  }{2\upsigma^2}} \right) \mathrm{d}I,x \leq \mathrm{SP},  \\
            \\
  \int_{-\infty}^{\mathrm{SP}} \frac{1}{\sqrt{2\uppi} \upsigma} \exp\left({-\frac { (I- \upmu)^2  }{2\upsigma^2}} \right) \mathrm{d}I  \\
 + \int_{\mathrm{SP}}^{x}  \uplambda\exp\left( -\uplambda\left(I-\rm SP\right) \right)  \mathrm{d}I ,x> \mathrm{SP}, 
             \end{array}
\right. \label{equ:gaussianexpon}
\end{equation}
where  $I$ is the observed \planck\ intensity.

 The four parameters are estimated by maximising this likelihood, and the noise and signal level are subsequently determined according to the two distributions.  As demonstrated in Figure~\ref{fig:snlevel}, we used a noise level of Gaussian mean $\upmu$, and a signal level of exponential median $\left(\rm SP +\ln\left(2\right)/\uplambda\right )$.

 \begin{figure*}
   \centering
 \includegraphics[width=0.98\textwidth]{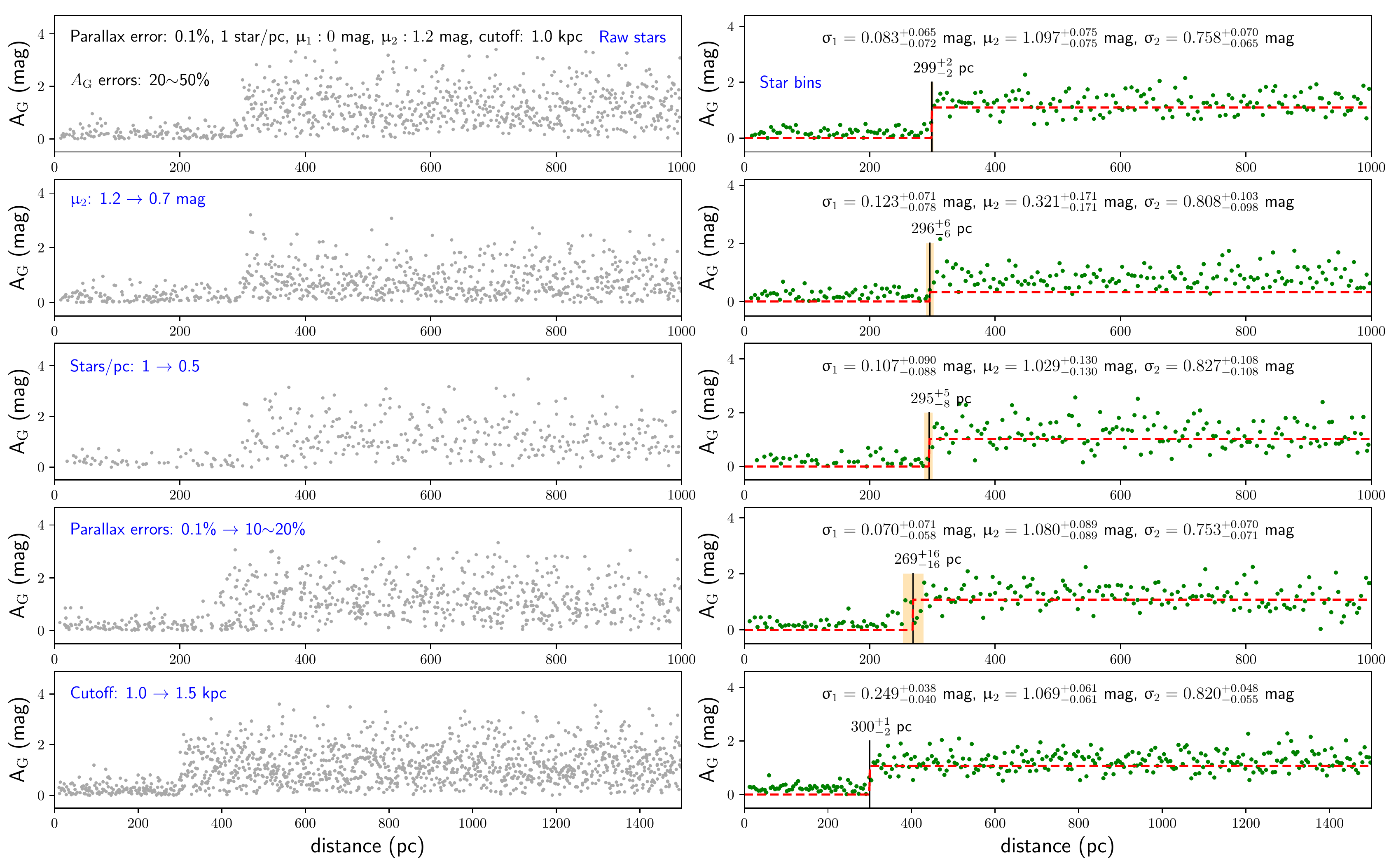} 
\caption{  Test of the model for calculating molecular  cloud distances with simulated on-cloud stars. The simulated molecular cloud is assumed to be located at a distance of 300 pc. The reference values of parameters are displayed in the top-left panel and we change one parameter at any one time to see the variation in  distance. The panels on  the left  column are the simulated raw data, while those on the right show the corresponding binned data and derived   distance results with the model. The raw extinction \ag\ and distances are averaged every 5 pc, weighted by their errors.  Only raw data are used in the distance estimation, while the binned data are only used to confirm the results by eye.  \label{fig:testmcmc}}
\end{figure*}

 There is a trade-off in the choice of the signal level. Lower signal levels keep more stars, which are good for statistical analysis, but involve many stars having low  extinction \ag,   smearing the breakpoints. Higher signal levels make the breakpoints evident but fewer stars remain. The signal level, the median value of the exponential component, is a compromised choice.  However,  in \S\ref{sec:choice}, we show that as long as the breakpoints are detected,   derived distances are insensitive to the choice of signal levels.

The signal and noise levels  classify on- and off-cloud stars, respectively. Although only on-cloud stars are employed to calculate distances, the off-cloud stars are useful as a reference to confirm the breakpoints.

\begin{figure*}[ht!]
   \centering
 \subfloat[]{\includegraphics[width=0.9\textwidth]{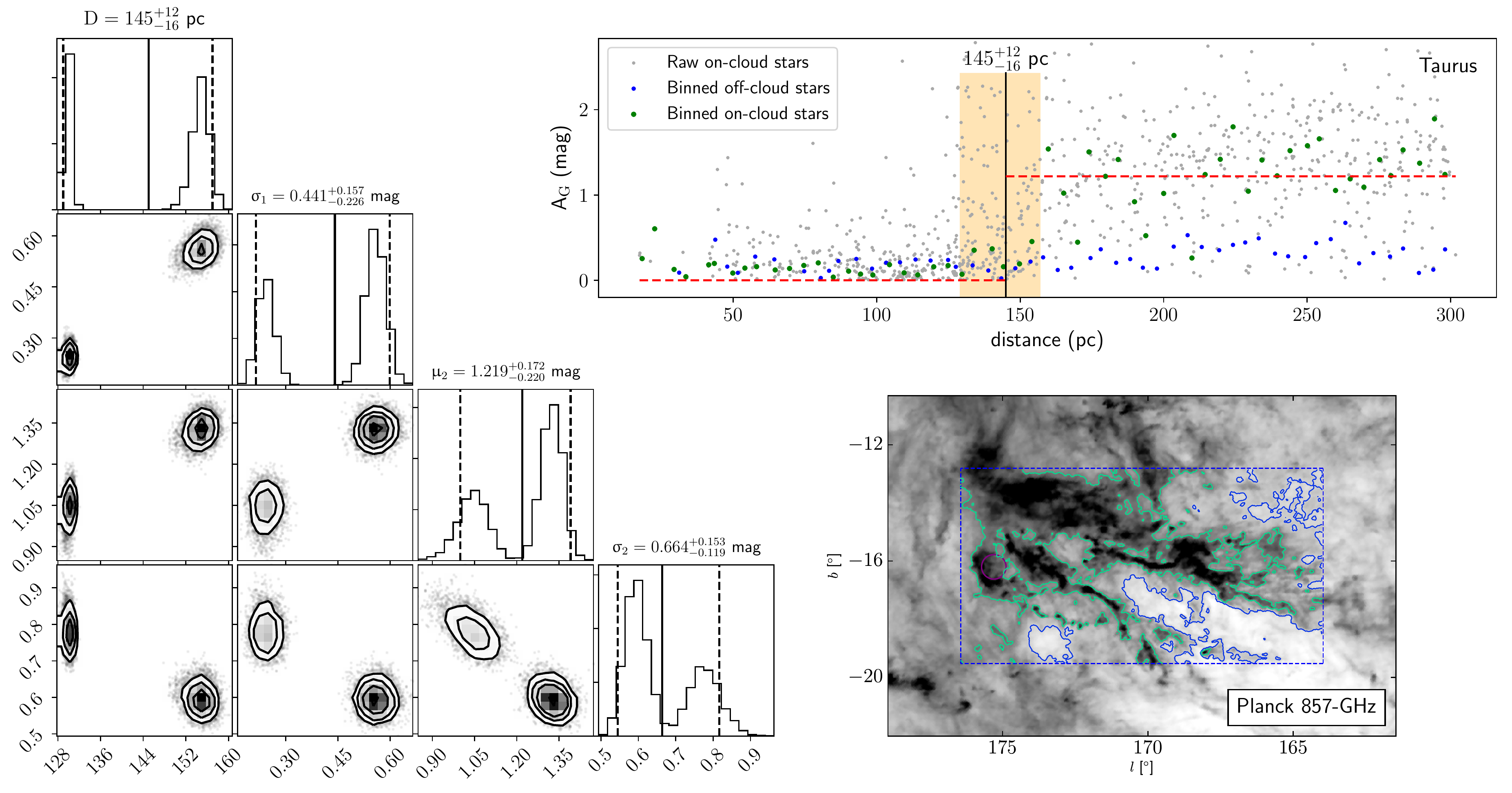}} \\
  \subfloat[]{\includegraphics[width=0.9\textwidth]{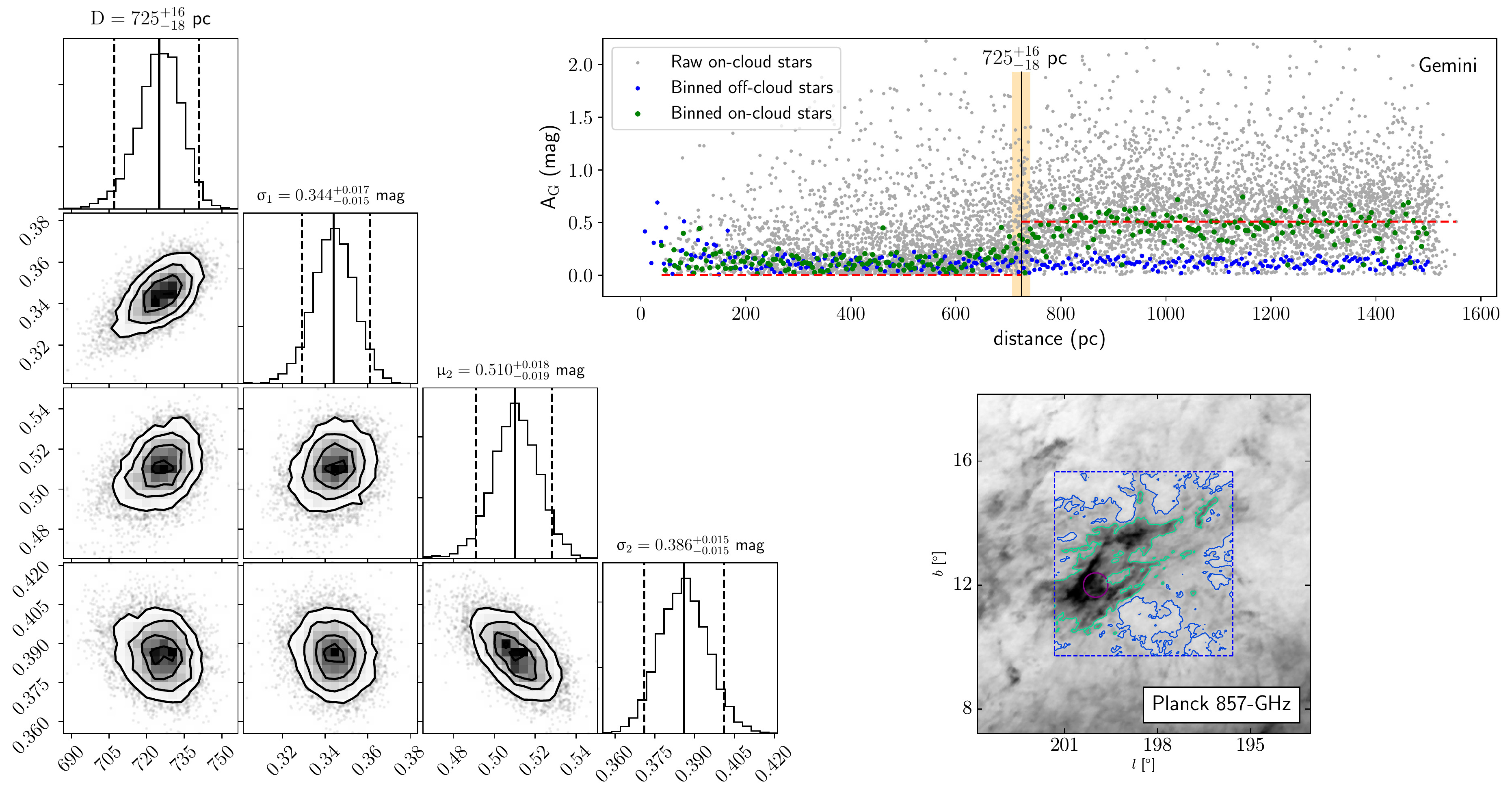}}
\caption{   The distance of Taurus (a) and  Gemini (b) molecular clouds.  In the panels on the bottom right, showing the \planck\ images, purple circles mark the position of molecular clouds, while  the blue and green contours correspond to the the noise and signal thresholds in Figure 2, respectively.    In the top-right panels,   the green and blue points present  on- and off-cloud stars (binned every 5 pc), respectively.    The dashed red lines are the modeled extinction \ag. The distances were derived with  raw on-cloud  \textit{Gaia} DR2 stars, which are represented with gray points. The black vertical lines indicate  the distance ($\rm D$) estimated with Bayesian analyses and MCMC sampling, and the shadow areas depict the  95\% HPD range of   distances.    The corner plots of the MCMC samples are displayed on the left. The mean and 95\% HPD of the samples  are shown with solid and dashed vertical lines, respectively,  and the systematic uncertainty is not included}. The Taurus molecular cloud  may contain two components.  \label{fig:tauDis} 
\end{figure*}

      \begin{figure*}[ht!]
 \centering
 \subfloat[]{\includegraphics[width=0.9\textwidth]{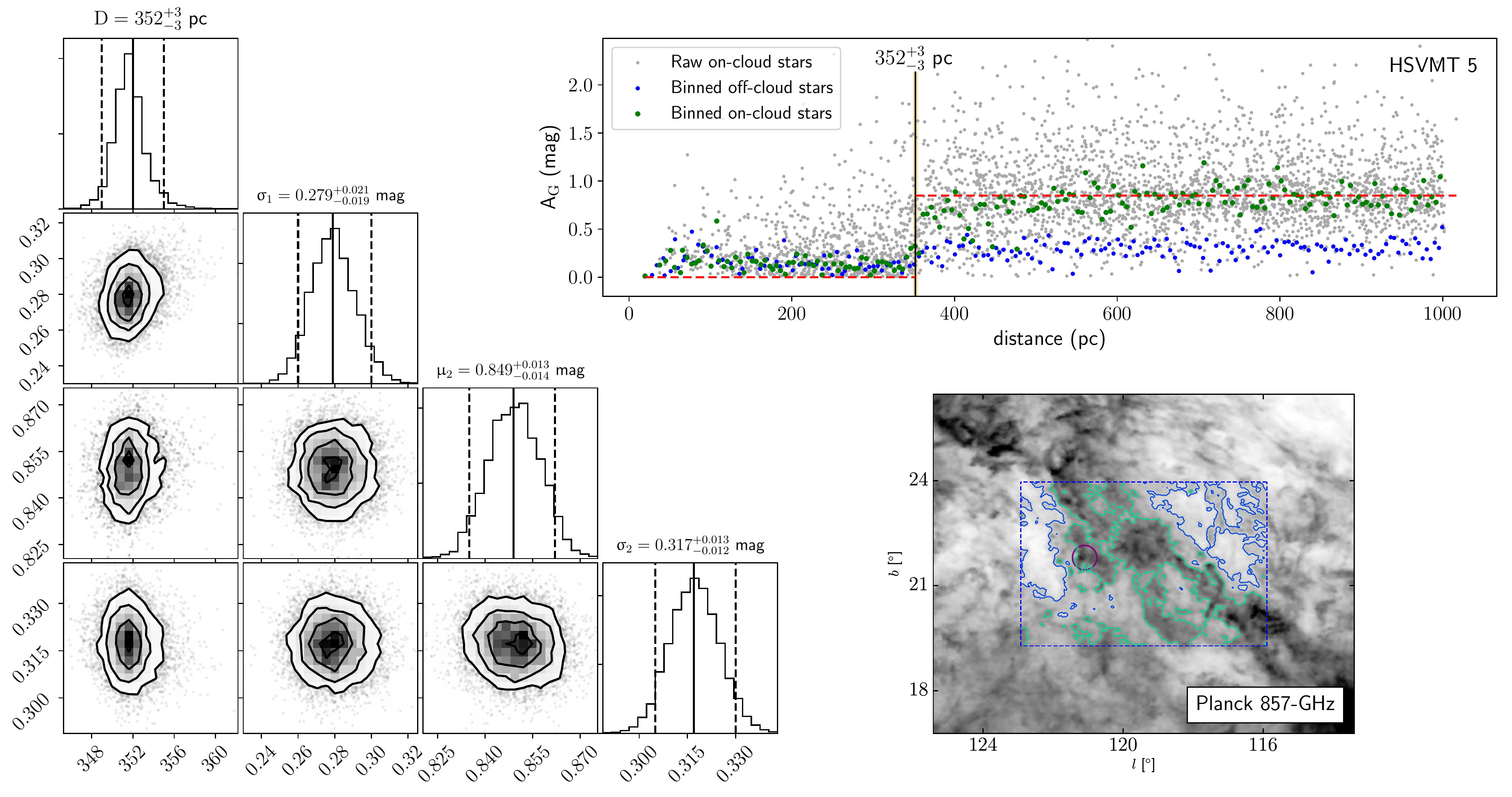}} \\
  \subfloat[]{\includegraphics[width=0.9\textwidth]{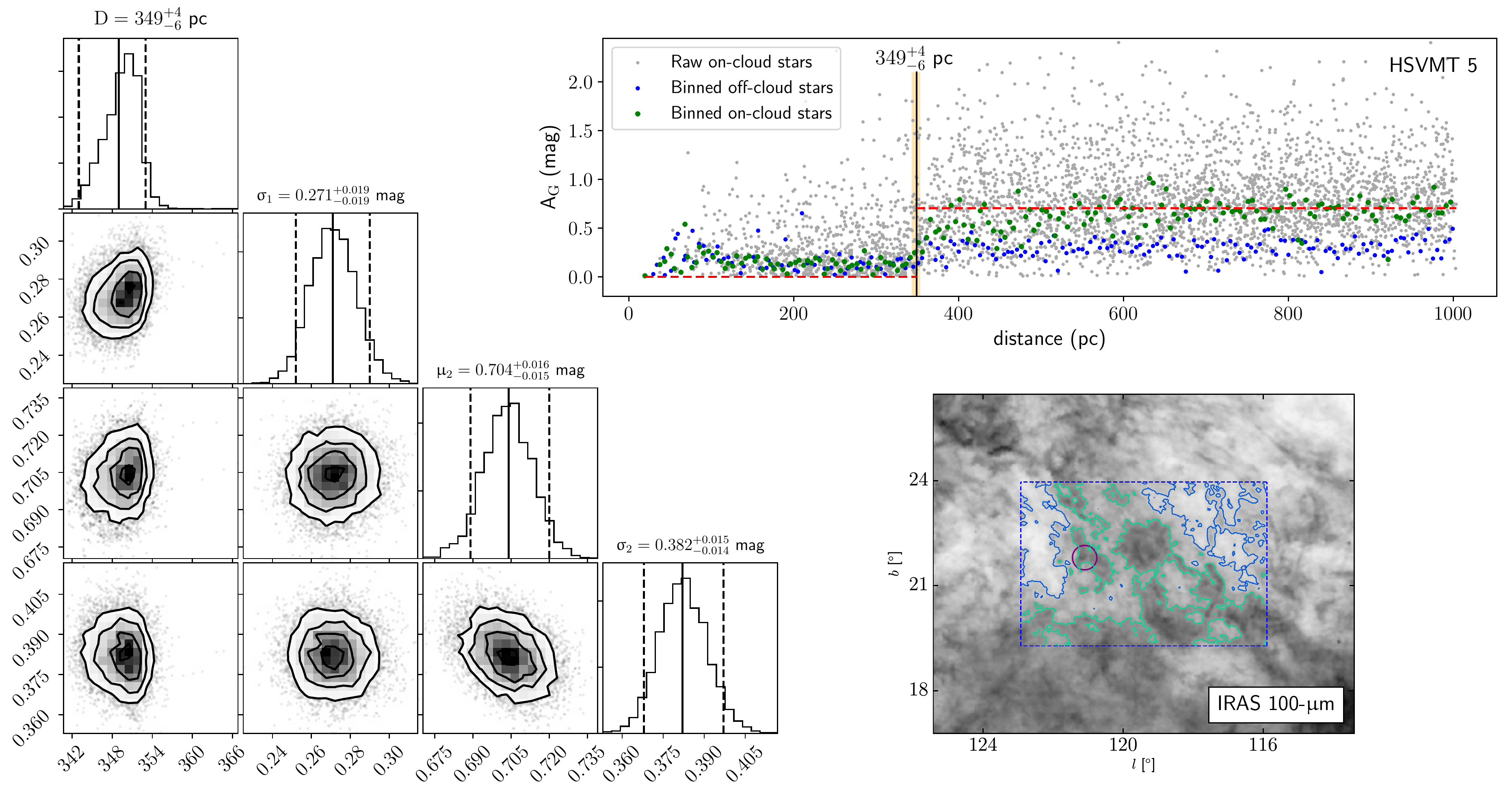}}

\caption{The distances  of molecular cloud HSVMT 5 produced with  \planck\ (a) and \irasum\ (b).  See the caption of Figure \ref{fig:tauDis}  for other details.  \label{fig:lhsvmt5Dis}}
\end{figure*}

 \subsection{ Estimating the distance}\label{sub:estimateDis}
 With the extinction  \ag\ and their corresponding distances of on- and off-cloud stars, we are now in the position to calculate the distance of molecular clouds.  We built a Bayesian model to estimate the distance with on-cloud stars  and solved for the parameters in the model with MCMC sampling. We emphasize that off-cloud stars are not involved in the model but only used to confirm the breakpoint by eye.


 The extinction \ag\ of on-cloud stars includes three components: (1) the foreground \ag, (2) the background \ag, and (3) the transition values. When passing through a molecular cloud along the line of sight, the extinction \ag\ gradually increases from the foreground \ag\ (which is usually 0 mag) level to the background \ag\ level. Here, we ignored the third component and focused on the average distance in our model.  Because the  minimum ($A_{\mathrm{G}}^{\mathrm{min}}$) and maximum  ($A_{\mathrm{G}}^{\mathrm{max}}$) values of the extinction \ag\ is 0 and 3.609 mag, respectively, we follow the procedure of \citet{2018A&A...616A...8A} using a truncated Gaussian distribution to model the  on-cloud star extinction \ag. 
 
 Our model involves  four  parameters, the cloud distance ($\rm D$), the extinction \ag\ dispersion ($\upsigma_1$) of foreground stars, and the extinction \ag\ ($\upmu_2$) and its dispersion ($\upsigma_2$) of background stars.  We found the extinction \ag\  of  foreground stars  ($\upmu_1$) is precisely zero but for only four molecular clouds, so we use  $\upmu_1=0$ in most cases. These four molecular clouds have additional small molecular cloud components in front of them, so  they will be treated particularly.  We used a  lower distance cutoff to remove the foreground stars of the small molecular cloud components, and their $\upmu_1$ were also modeled, i.e., five parameters in total for them. 

 With those four parameters and  a list of distances $\mathrm{d}_i$ with   standard deviations  $\Delta \mathrm{d}_i$, which were calculated with the  reciprocal of 10000  parallax samples (see \S\ref{sec:data}),  and extinction   $A_{\mathrm{G}i}$ (with standard deviations   $\Delta A_{\mathrm{G}i}$), we derive an appropriate likelihood below. This approach requires no binning, and the  binned extinction \ag\ and distances are only used for eye confirmation. 

We calculated the likelihood of each star on the condition of being in front of or behind the molecular cloud. Denoting the CDF of the standard normal distribution as  
\begin{equation}
\upphi\left(x\right)=  \frac{1}{\sqrt{2\uppi}} \int_\infty^x e^{-t^2/2} \mathrm{d}t 
\end{equation}
 and given $\rm D$, the probability for a star to be in front of  the molecular cloud is  
\begin{equation}
f_i=\upphi\left( \frac{\rm D-\mathrm{d}_i }{ \Delta \mathrm{d}_i }  \right),
\end{equation}
while the probability to be located behind the molecular cloud is $1-f_i$.

The form of  a truncated Gaussian distribution \citep{2018A&A...616A...8A}, with mean $\upmu$ and  standard deviation  $\upsigma$, is
\begin{equation}
p\left(A_{\mathrm{G}i} |  \upmu, \upsigma\right) =  \left\{
             \begin{array}{ll}
          \frac{ \frac{1}{ \upsigma\sqrt{2\uppi}} \exp\left( -\frac{1}{2}\left( \frac{A_{\mathrm{G}i}-  \upmu}{  \upsigma}    \right)^2 \right) }{  \frac{1}{2}\left( \mathrm{erf}\left(\frac{ A_{\mathrm{G}}^{\mathrm{max}}  - \upmu}{  \sqrt{2}\upsigma}  \right) +  \mathrm{erf}\left(\frac{ \upmu- A_{\mathrm{G}}^{\mathrm{min}}   }{  \sqrt{2}\upsigma } \right)   \right) } , &A_{\mathrm{G}}^{\mathrm{min}}\leq A_{\mathrm{G}i}  \leq A_{\mathrm{G}}^{\mathrm{max}},\\
          & \\
 0,  & \mathrm{otherwise}, 
            \end{array}
\right. \label{equ:model}
\end{equation}
 where 
\begin{equation}
\mathrm{erf} \left( z \right)=\frac{2}{\sqrt{\uppi}} \int_0^z e^{-t^2} \mathrm{d}t.
\end{equation}

 In order to consider the measure error of the extinction \ag, we convolved the   standard deviation  of   the extinction \ag, $\Delta A_{\mathrm{G}i}$, with $\upsigma_1$ and $\upsigma_2$. Consequently, the likelihood  of foreground stars is 
\begin{equation}
PF_{i}=p\left(A_{\mathrm{G}i} |  \upmu_1, \sqrt{\upsigma_1^2+\Delta A_{\mathrm{G}i}^2} \right).
\label{equ:pf}
\end{equation}
Similarly, the likelihood of background stars is
\begin{equation}
PB_{i}=p\left(A_{\mathrm{G}i} |  \upmu_2, \sqrt{\upsigma_2^2+\Delta A_{\mathrm{G}i}^2} \right).
\label{equ:pb}
\end{equation} 
 Consequently, the likelihood of a star is 
\begin{equation}
p\left(A_{\mathrm{G}i} |  \upmu_1, \upsigma_1 ,\upmu_2, \upsigma_2,\rm D\right)=   f_i  PF_{i}+\left(1-f_i\right)PB_{i},
\label{equ:singlelike}
\end{equation}

The total likelihood is the product of the  likelihoods over all the stars, and we solve this model with MCMC sampling.  In order to obtain a high sampling rate, we may need to  set smart priors.  The prior distribution of $\rm D$ is assumed to be uniform, so that the model can uniformly search the switch point, i.e., the breakpoint of the extinction \ag, along the line of sight.     Denoting the minimum and  maximum of $\mathrm{d}_i$ as $\rm D_{\rm min}$ and  $\rm D_{\rm max}$, the prior of $\rm D$ is $\mathcal{U}\left(\rm D_{\rm min},\rm D_{\rm max} -50 \ pc \right)$, where $\mathcal{U}$ represents the uniform distribution and the 50 pc is set to avoid touching the edge. In practice, instead of sampling $\upsigma_1$ and $\upsigma_2$, we sampled their reciprocal, denoted as $\rm I\upsigma_1$ and $\rm I\upsigma_2$. The prior distributions of    $\rm I\upsigma_1$,  $\upmu_2$, and   $\rm I\upsigma_2$  are assumed to be exponential, and here, we use a form of $\mathcal{E} \left(\beta\right)$ for the exponential distribution, where $\beta$ is the mean.  We summarize the priors as 
\begin{equation}
\left\{
             \begin{array}{rcl}
  \rm D   & \sim &\mathcal{U}\left(\rm D_{\rm min},\rm D_{\rm max}-50\ pc  \right), \\
               &&\\
  \rm I\upsigma_1  & \sim &\mathcal{E}\left(2\right), \\
  &&\\
\upmu_2  &\sim& \mathcal{E}\left(\upmu_{50}\right),\\
  &&\\
 \rm I\upsigma_2  & \sim &\mathcal{E}\left(\rm I\upsigma_{50}\right), 
             \end{array}
\right.
\end{equation} 
where   $\mathcal{U}$  and  $\mathcal{E}$ represent  the uniform and exponential distributions, respectively,   $\upmu_{50}$ and $\rm I\upsigma_{50}$ are the mean and reciprocal standard deviation  of extinction \ag\  of stars with distances $>$$\left(\rm D_{\rm max}-50\ pc\right)$, and the initial guess of 2 mag$^{-1}$ for $\rm I\upsigma_1 $ is derived from the reciprocal of 0.45 mag, which is the typical extinction \ag\    standard deviation  of clustering stars derived by \citet{2018A&A...616A...8A}.  

In order to decrease the autocorrelation time of the MCMC samples, we  thinned the samples. Because the acceptance rate is about 25\%,   we thinned the samples by a factor of 15,  considering only  every 15th step in each chain, and the autocorrelation time is small (about 4) after thinning. We calculated eight independent chains, and each chain has 1000   thinned samples (with extra 50  burn-in, which means the first 50   thinned samples were removed), i.e.,  8000  thinned  posterior samples for each parameter. In order to increase the sampling speed, we used the Gibbs sampler  \citep{geman1984stochastic}, i.e., we changed one parameter at a time and  used Metropolis–Hastings algorithm \citep{metropolis1953equation,hastings1970monte} for each time until all four parameters have obtained the next values. The state transition function of the parameters are all Gaussian, whose  standard deviations   are 50 pc for $\rm D$, 0.3 mag for $\upmu_2 $, and 0.3 mag$^{-1}$ for $ \rm I\upsigma_1$ and $ \rm I\upsigma_2$. 

We investigated the convergence of each chain with the Gelman–Rubin diagnostic \citep{gelman1992inference}. The potential scale reduction factor (PSRF), $\hat R$, is less than 1.01 for all molecular clouds but Taurus (whose $\hat R$ is about 6), which has two close components. Consequently, the MCMC chains  have converged but for Taurus.

 In order to verify  the MCMC results, we compared the Gibbs sampler with the affine-invariant ensemble sampler \citep{2010CAMCS...5...65G}  implemented by  the Python MCMC module EMCEE \citep{2013PASP..125..306F}. The EMCEE package usually produces many outliers, but when it occasionally generates less outliers,  it gave the same results with the Gibbs sampler.

\begin{figure*}
   \centering
 \includegraphics[width=0.9\textwidth]{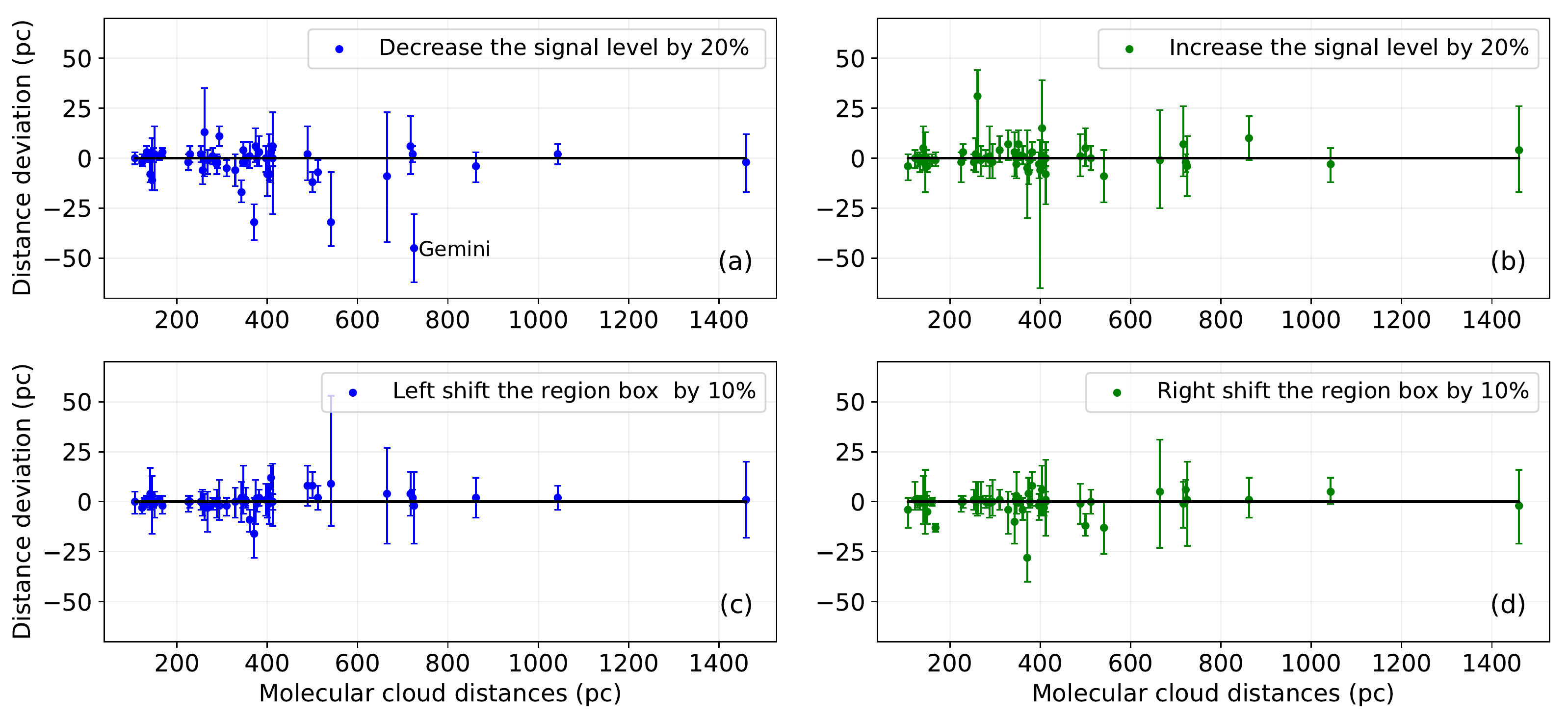} 
\caption{Distance discrepancies and errors  with altered parameters.  The effect of molecular cloud region boxes is shown in (a) and (b) and the signal levels in   (c) and (d).   Only one  molecular cloud, Gemini, shows large discrepancies (with magnitudes $>$40 pc), which is  in case (a).  \label{fig:comparebin}}
\end{figure*}

In order to balance  the star numbers  of the two truncated Gaussian components in Equation \ref{equ:model} and to avoid the interference of farther molecular clouds along the line of sight, we set a distance cutoff for each molecular cloud. Because we need to know the distance before setting the distance cutoff, this is a recursive process.  Usually, a distance cutoff of 1000 pc is sufficient, and we adjust this cutoff for  far or near molecular clouds if necessary. Close molecular clouds usually have too few foreground stars compared to background stars, and small fluctuation of \ag\ in the  background stars would make their distances wrongly recognized.

A lower cutoff for foreground stars  is unnecessary but for four molecular clouds,  which are the far component of Cepheus,  Mon R2, Polaris, and  Rosette. In oder to  remove the effect of foreground components, we removed on-cloud stars  that are nearer than  400,  500, 325, and 1000 pc for them and the corresponding $\upmu_1$  are estimated to be  0.683,  0.499, 0.295,  and 0.839 mag, respectively.  Because of having 5 parameters to model, we calculated 10 chains for these four molecular clouds, i.e., 10000 thinned samples for each of them. 
 
 \subsection{ Testing the model}\label{sub:testmodel}

We tested this model of calculating molecular cloud distances with simulated data before applying it to \textit{Gaia} DR2 stars.  The simulated a molecular cloud  is at a distance of  $\rm D=300$ pc, and its   foreground star extinction is 0 mag, i.e., $\upmu_1=0$. The dispersions of the foreground and background stars \ag\ are 0.3 and 0.6 mag, respectively, i.e., $\upsigma_1=0.3$ mag and $\upsigma_2=0.6$ mag. We added scatter \ag\ errors of 20$\sim$50\% on the extinction \ag, and stars whose extinction \ag\ $<$0 or $>$ 3.609 mag were removed.
 
 We changed four parameters, the background star extinction $\upmu_2$, number of stars per pc, the relative parallax errors, and the distance cutoff, to see the variation on the resulting distances. As demonstrated in Figure~\ref{fig:testmcmc}, our model detected the distances successfully in all cases. Unsurprisingly,  smaller jumps of  extinction \ag, fewer \textit{Gaia} DR2 star samples, and larger parallax errors would cause larger distance errors, while the distance cutoff has no effect in the distance determination. Remarkably, although the distance errors became larger, the distance was still consistently recognized. 

In \textit{Gaia} DR2, the errors on distances are usually better than 10\% for stars nearer than 1 kpc. Therefore, the two most  important factors that may affect our results are the number of on-cloud stars and the  magnitude of \ag\ caused by  molecular clouds.

 \subsection{The distance of  the Taurus and Gemini molecular clouds}\label{sub:testmodel}

 We applied our model to  the Taurus and Gemini molecular clouds. The corner plots of the four  parameters, $\rm D$, $\upsigma_1$,  $\upmu_2$, and $\upsigma_2$, are displayed in Figure~\ref{fig:tauDis}, together with their means and 95\%  Highest Posterior Densities (HPDs). On the bottom right, the dashed blue boxes of (a) and (b) are the manually chosen cloud regions (see \S \ref{sub:snlevel}), which contain molecular clouds and extra noise regions. The blue and green contours represent the noise and signal thresholds, respectively, which were  determined in \S\ref{sub:snlevel} (see Figure~\ref{fig:snlevel}).  On the top right, we display the   distances and on- and off-clouds stars with green and blue points, respectively. The derived distances to the Taurus and Gemini molecular clouds are $145_{-16 }^{+12 }$ and $  725_{-18 }^{+16 }$   pc, respectively. 

 Interestingly, the Taurus molecular cloud seems to have two components.  Farther than 130 pc, there are still many stars having low extinction \ag\  until about 150 pc. The Taurus  distance given by \citet{2009ApJ...698..242T} is 161.2$\pm$0.9 pc, which may be corresponding to this component.

 \section{Molecular clouds distances}
 \label{sec:distance}
 
 \subsection{\planck\ versus IRAS 100-$\mu$m data}\label{sec:planckvsiras}

We demonstrate that  \planck\ is a better dust tracer than  \irasum; or at least is more well-suited to our method.   As an example, we show the distances of the molecular cloud HSVMT 5 produced with \planck\ and \irasum\ in Figure ~\ref{fig:lhsvmt5Dis}.
 
The distance estimated with \planck\   is $ 352_{-3}^{+4}$  pc, and $  349_{-6  }^{+4  }$ pc for \irasum.  As indicated in Figure \ref{fig:lhsvmt5Dis}, the average extinction of on-cloud stars classified with \planck\ is higher than that with \irasum,  suggesting that the on-cloud stars identified with  \planck\ is less contaminated by other \textit{Gaia} stars that have low extinction \ag. Consequently, the extinction values, \ag, of  the background stars are better separated from the  foreground stars with \planck\ emission.

\begin{table} 
\centering
\caption{Comparing with VLBI distance measurements.} \label{tab:comorion}
\begin{tabular}{ccccc}
 
\hline
\hline
Cloud &    \textit{Gaia} DR2 &    VLBI &References  \\
  &    (pc)&    (pc)   &(pc)  \\
\hline 
L1641 &   $  408_{-4  }^{+4  }$ &    428$\pm$10 & 1 \\
NGC 2068 &  $  412_{-4  }^{+4  }$   & 388$\pm$10  &   1  \\
Mon R2&  $  862_{-10 }^{+10 }$    & $ 893_{-40}^{+44}$  & 2  \\
Ophiuchus& $  131_{-2  }^{+2  }$   & 137.3$\pm$1.2  & 3 \\
Perseus& $  310_{-4  }^{+4  }$   & 293$\pm$22  or 321$\pm$10 &  4  \\
\hline
\end{tabular}
\tablebib{
(1)~\citet{2017ApJ...834..142K};   (2)  \citet{2016ApJ...826..201D}; (3)  \citet{2017ApJ...834..141O}; (4) \citet{2018ApJ...865...73O}. In Perseus, \citet{2018ApJ...865...73O} obtained a distance of  321$\pm$10 pc (IC 348)    with VLBI, while the distance of NGC 1333 is 293+/-22 pc suggested by young stars in Gaia DR2.}
\end{table}
 
\subsection{The impact of the parameter choice}
\label{sec:choice}

In this subsection, we examine the reliability of our method  and its dependence on parameters. In our method, we primarily involve two  subjective parameters: the manually chosen molecular cloud region  and   the signal level  cutoff in the \planck\ intensity.  In total, we found  52 molecular clouds have their distances well determined with the chosen molecular cloud regions and signal levels. 

In order to see the influence of the  two parameters, we shifted them one at a time and compared the distance variations. In Figure \ref{fig:comparebin}, we display the distance discrepancies and uncertainties after altering parameters. To reveal the effect of the molecular cloud region box, we shifted the molecular cloud region box along $l$ to the left and right by 10\% of the box size in $l$, while  the signal levels were altered by $\pm$20\%. The distances show slight fluctuations in the molecular clouds, while their uncertainties widely agree with the distances calculated with the chosen parameters. Clearly, large distance discrepancies usually yield large uncertainties and no evident  systematic deviations are present and only Gemini  shows large distance deviations (with magnitudes $>$40 pc) in case (a), confirming our claim that correct recolonization of  breakpoints guarantees comparable distances.

Consequently, our distance estimation is robust and only weakly dependent on the choice of  region boxes  and signal levels in a reasonable range. In addition, in order to make sure the breakpoints are genuine,  it is necessary to confirm the breakpoints with off-cloud stars.

 \subsection{The Distance catalog}

With the method described in \S\ref{sec:method}, we calculated distances for many  high-Galactic-latitude molecular clouds, most of which have been cataloged by \citet{1985ApJ...295..402M},  \citet{1996ApJS..106..447M}, and \citet{2014ApJ...786...29S}.  In these three catalogs, many molecular  clouds belong to the same molecular could complexes or filaments, and we only give one distance for each complex or filament, because  of  the large dispersion in the extinction \ag\ and inadequate \textit{Gaia} stars when dealing with small subregions.

  \begin{figure}
   \centering
 \includegraphics[width=0.48\textwidth]{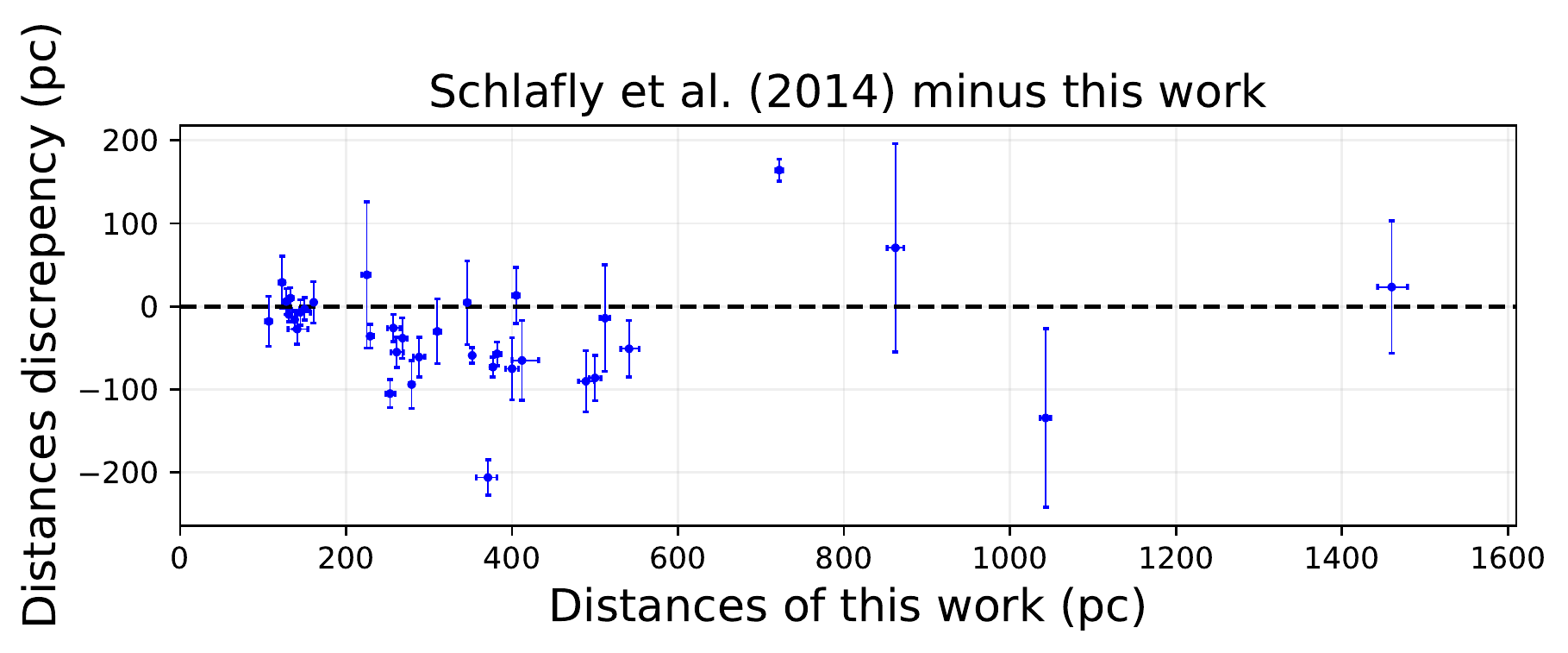} 
\caption{A comparison of distances in our work with that from \citet{2014ApJ...786...29S}.   \label{fig:compare}}
\end{figure}
 
We have reproduced the distances of many star-forming regions, whose distances are  well determined with  VLBI measurements, such as Orion \citep{2017ApJ...834..142K}, Mon R2~\citep{2016ApJ...826..201D},  Ophiuchus \citep{2017ApJ...834..141O}, and Perseus \citep{2018ApJ...865...73O}. In Table \ref{tab:comorion}, we compare our results with VLBI-measured distances.  Considering the distance errors and dispersions, the   molecular cloud distance given by \textit{Gaia} DR2 agrees  quite  well with VLBI measurements.  We suggest that VLBI and \textit{Gaia} DR2 may see different components of the molecular  clouds, and  the slight discrepancies  may be due to the structure of the molecular clouds.

We summarize the distances in Table~\ref{tab:distance},   listing 52  molecular clouds  whose distances are well determined, i.e., the breakpoints are evident. Many molecular clouds cannot have their distance determined because (1) they are not well defined in \planck, (2) their  optical depths are too low, or (3) their covering areas are too small.      In the second last column, we mark those molecular clouds whose distances are not provided by \citet{2014ApJ...786...29S} or measured using VLBI  with  ``Y'', 13  in  total.

The distance figures of all molecular clouds have been provided on Harvard Dataverse\footnote{\url{https://doi.org/10.7910/DVN/C6YO4T}}.

   \begin{figure*}[ht!]
 
 \centering
 \subfloat[]{\includegraphics[width=0.9\textwidth]{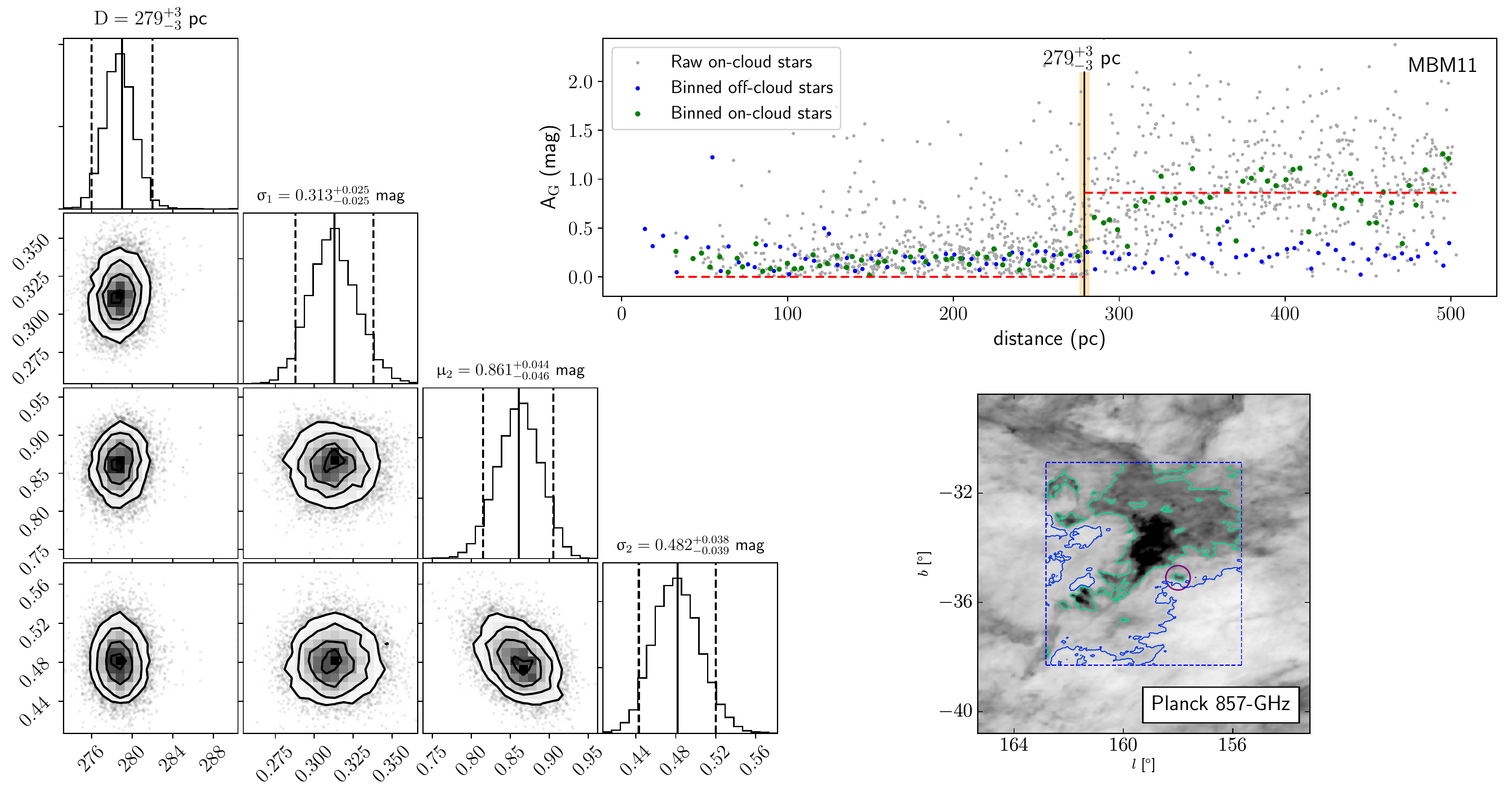}} \\
  \subfloat[]{\includegraphics[width=0.9\textwidth]{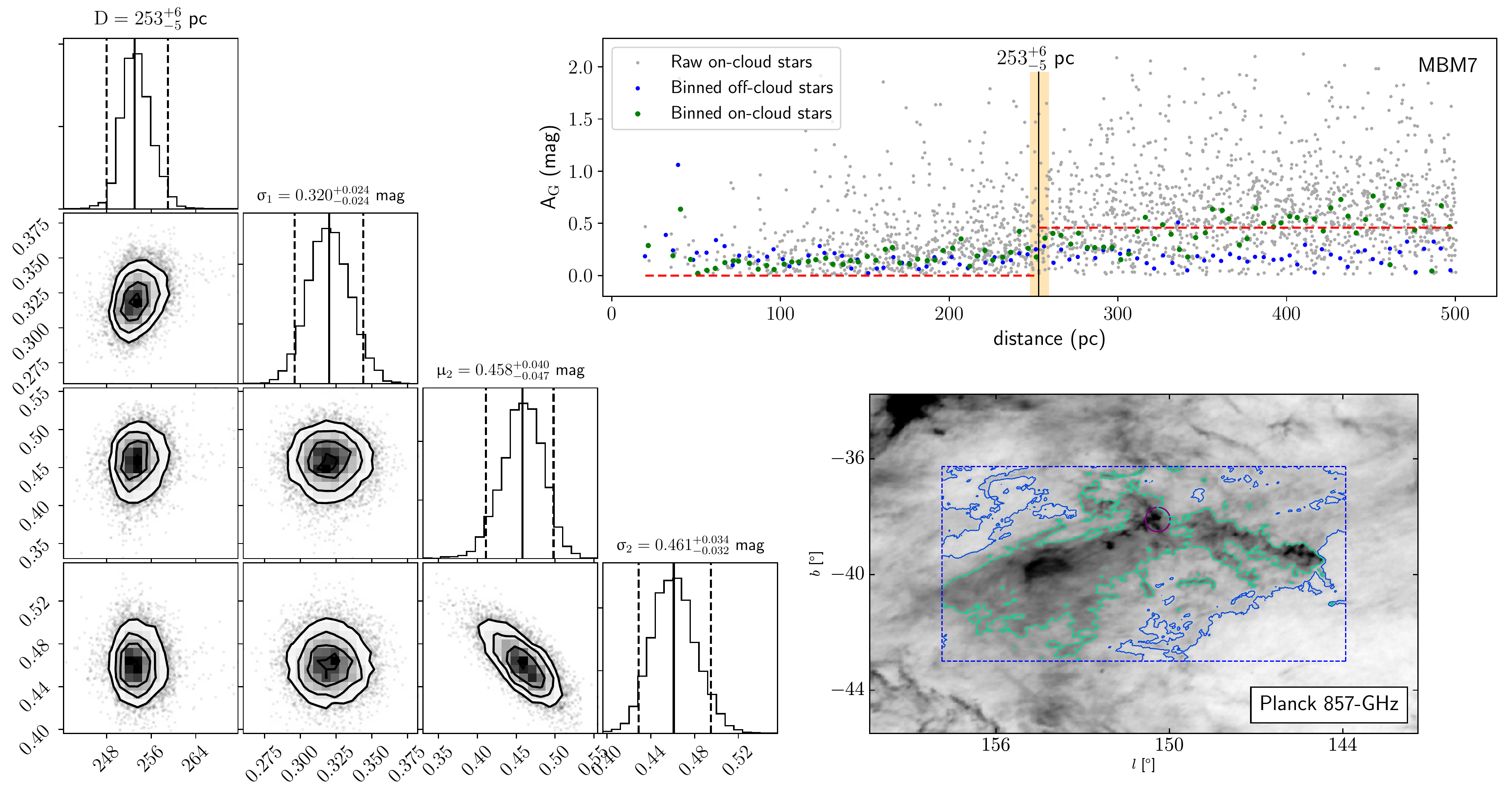}}
     
\caption{The distance of MBM11 (a) and MBM7 (b). These two molecular clouds show large distance discrepancies with that provided by \citet{2014ApJ...786...29S}.   See the caption of Figure \ref{fig:tauDis}  for other details.  \label{fig:mbm1117Dis}}
\end{figure*}

\subsection{ The systematic error}

 In this section, we estimate the systematic error in the distances.

 First, as suggested by the testing results (see Figure \ref{fig:testmcmc}),  the distances show a systematic error of 1\%-10\%. In the testing data, we simulated parallax errors of 10\%-20\%, and within 1 kpc from the Sun, however the uncertainties of Gaia DR2 parallaxes are less than 10\%. Therefore, the true systematic error is likely to be smaller. 

 Secondly, as shown in Figure~\ref{fig:comparebin}, the root-mean-square (RMS) of the relative distance deviations in the four cases are 3\% (a), 2\% (b), 1\% (c), and 2\% (d), respectively, suggesting a systematic error of about 3\% caused by subjective choices.

 Thirdly, assuming the VLBI results (see Table \ref{tab:comorion}) are the true distances, our distances show a systematic error of $\leq$ 6\%. However, VLBI usually measures the parallax of one single star, which may deviate from the main part of the molecular clouds. 

 Fourthly, \citet{2018A&A...616A...2L}   mentioned  that the Gaia DR2 parallaxes have a systematic error of approximately 0.1 mas. Considering a molecular cloud at a distance of 500 pc (2 mas), this parallax systematic error causes a systematic error of  about 5\% in the distance. 
 
 Consequently, we estimate that the systematic distance error  is about 5\%, possibly slightly larger for distant molecular clouds.

 \section{Discussion}

 \begin{figure*}[ht!]
 
 \centering
 \subfloat[]{\includegraphics[width=0.9\textwidth]{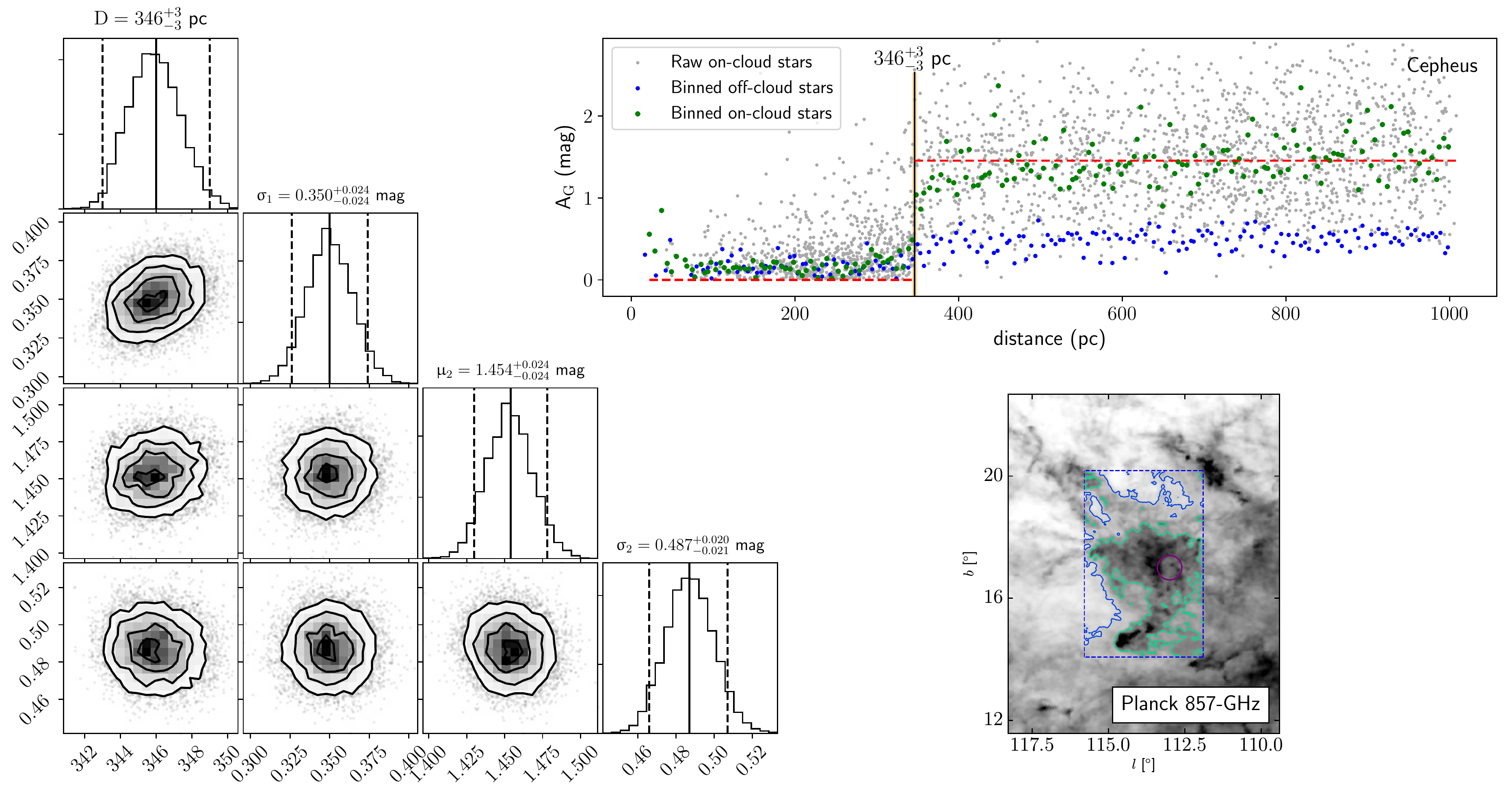}} \\
  \subfloat[]{\includegraphics[width=0.9\textwidth]{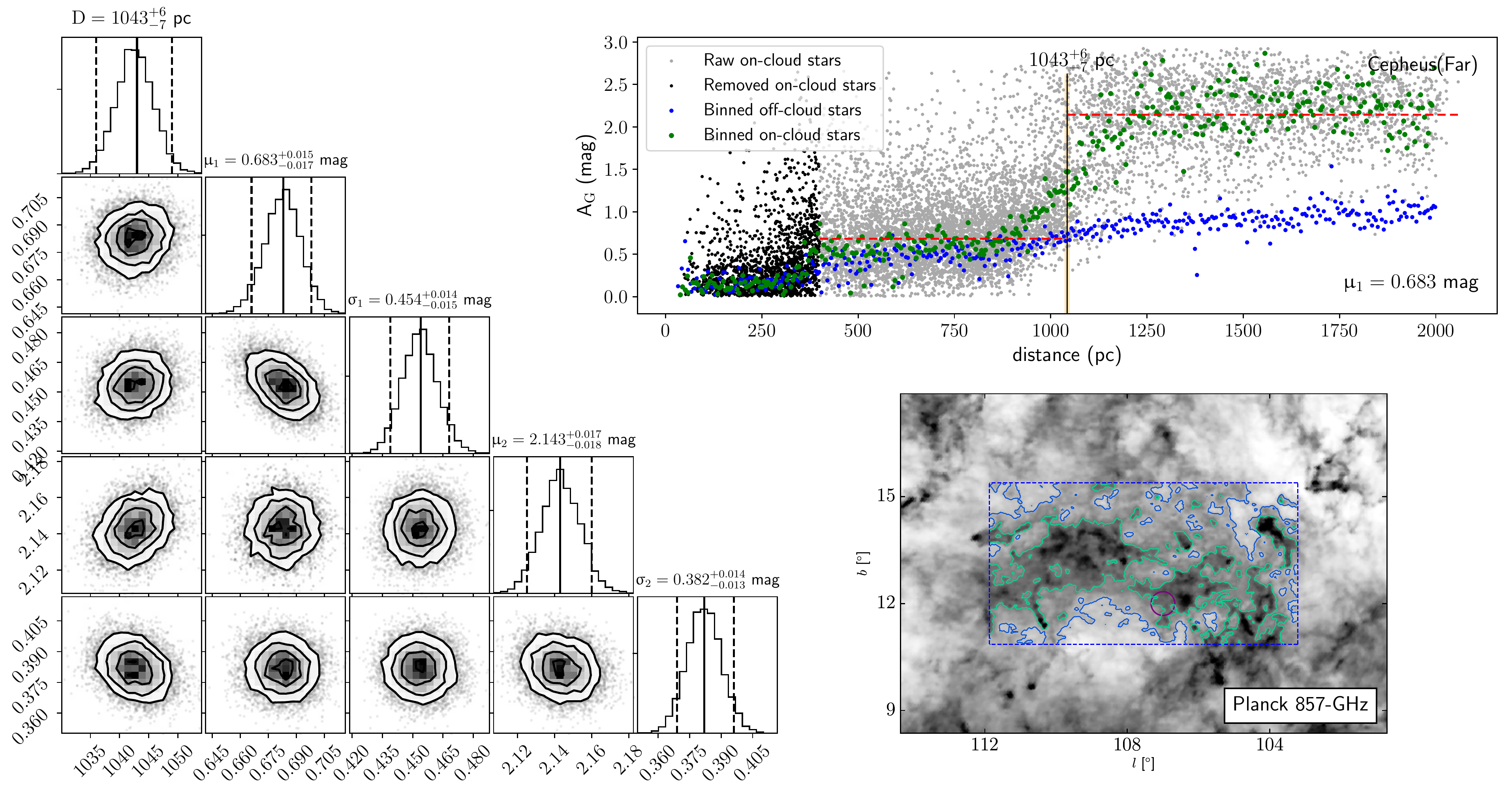}}
 
\caption{The distances of near (a) and far (b) components in the Cepheus molecular cloud.   See the caption of Figure \ref{fig:tauDis}  for other details.  \label{fig:CepheusDis}}
\end{figure*}

 \subsection{Comparison with previous results}

  In Figure~\ref{fig:compare}, we draw the distances derived from \textit{Gaia} DR2 against that provided by \citet{2014ApJ...786...29S}. We averaged the distances weighted by the square inverse of  their errors when multiple  distances (corresponding to the same molecular cloud) are provided by \citet{2014ApJ...786...29S}. The distances provided by \citet{2014ApJ...786...29S} display a slight systematic shift, $\sim$24 pc, towards the farther distances. In terms of relative errors, the systematic shift is, after removing one  outliers ($>$70\%),  about 13\%.

This systematic shift was predicted by \citet{2014ApJ...786...29S}. They attribute the systematic errors to stellar models, dust models, and the reddening law.  However, in addition to these,  we found the  molecular regions used by \citet{2014ApJ...786...29S} are much smaller than that in our work, which means their results may only represent the distances of molecular clouds in those small regions, and are less affected by the thickness of the molecular clouds along the line of sight. Alternatively, nearer clouds are brighter and easily to be chosen, which could cause the distances given by  \citet{2014ApJ...786...29S}   systematically  smaller.  

However, the \textit{Gaia} DR2 data may also  contribute to this systematic error. The  $\sim$0.1 mas systematic parallax error in \textit{Gaia} DR2 \citep{2018A&A...616A...2L} may be one of the causes of the systematic distance discrepancy. At a typical distance of 500 pc, 0.1 mas error corresponds to $\sim$25 pc shift, which can explain this systematic shift. Furthermore,  the extinction \ag, which has large uncertainties, may contain unknown systematic errors.

One means of examining  the discrepancy is to compare the distance modulus in \citet{2014ApJ...786...29S} with the \textit{Gaia} DR2 parallaxes. However, this is beyond the scope of this work, and alternatively, in Figure~\ref{fig:mbm1117Dis}, we display the distances of MBM11 and MBM7, which show large distance discrepancies. As shown in Figure~\ref{fig:mbm1117Dis}, the distances of MBM11 and MBM7 are  $  279_{-3  }^{+3  }$  and  $  253_{-5  }^{+6  }$    pc, respectively. However, in the distance catalog  of  \citet{2014ApJ...786...29S},   the  distance of MBM11 molecular cloud is approximately 206$\pm$23 pc, which is the average distance of MBM11, 12, 13, and 14, while the distance of MBM7 is  $148_{-11 }^{+13}$ pc. Based on \textit{Gaia} DR2, the distances of these two molecular clouds, particularly MBM11, are well determined.

\subsection{Individual molecular cloud distances}

In this subsection,  we discuss the distances of several  individual molecular clouds.   \citet{2014ApJ...786...29S} obtained a distance of about 350 pc for the Ursa Major molecular clouds, which is much farther than previous studies, $\sim$110 pc \citep{1993ApJS...88..433P}. Using \textit{Gaia} DR2 data, we have derived a distance of $  412_{-12 }^{+20 }$  for   the Ursa Major molecular clouds, which is close to  that given by \citet{2014ApJ...786...29S}.

 \citet{2015AJ....150...60L} performed a  survey of two CO isotopologue lines  towards the same  Gemini region as shown in Figure~\ref{fig:tauDis}, and they used a distance of 400 pc, but \textit{Gaia} DR2 clearly shows that its distance is   about 725 pc. Consequently, the masses  and sizes of Gemini molecular cores calculated by \citet{2015AJ....150...60L}  should be revised accordingly.

The Cepheus~\citep{1989ApJ...347..231G, 1997ApJS..110...21Y, 2009ApJS..185..198K} is an interesting region. As shown in Figure~\ref{fig:CepheusDis}, obviously, there are two components along the line of sight.  The distance of the nearer one is $  346_{-3  }^{+3  }$  pc, while the distance of the farther one is $ 1043_{-7  }^{+6  }$  pc.  According to the CO observations of \citet{1989ApJ...347..231G}, the radial velocity range of the far component is  about -12  \kms, while the nearer component has a velocity of about 0 \kms. The distances of the two components derived by \citet{1989ApJ...347..231G} are $\sim$300 and 800 pc, respectively, which are consistent with our results.



 

 \section{Summary}
 
 Using  the parallaxes and   extinction \ag\ provided by \textit{Gaia} DR2, we derived  the distances for 52 molecular clouds, most of which are at high Galactic latitudes, i.e., $|b|>10$\deg.  The systematic error of the distances is about 5\%, and 13 molecular clouds have reliable distances determined for the first time,   In addition, we have confirmed the distances of many star-forming regions, such as Orion, Taurus,   Cepheus,  and Mon R2.  

We used  \planck\ data  rather than CO data to trace molecular clouds because CO observations are incomplete at high Galactic latitudes. However, at low Galactic latitudes  ($|b|\leq10^{\circ}$),  multi-wavelength observations with high spatial resolutions would classify on-  and off-cloud stars   more accurately. 

 \textit{Gaia} DR2  has enabled us to determine the distances of  many nearby molecular clouds efficiently. Although the  large errors of extinction \ag\ in \textit{Gaia} DR2 prevent us from examining distant ($>$ 2 kpc) molecular clouds, with the improved qualities of distances and extinction \ag\ in future \textit{Gaia} data releases,  we would be able to obtain the distances of many more molecular clouds.

 \begin{acknowledgements}
 We thank Mario G. Lattanzi for his valuable comments and careful proofreading. We are grateful to an anonymous referee for his/her constructive comments, particularly on the distance likelihood and the MCMC method. This work was sponsored by the Ministry of Science and Technology (MOST) Grant No.  2017YFA0402701,   the 100 Talents Project of
the Chinese Academy of Sciences (CAS),  the National Science
Foundation of China under Grand NO. 11673051 and 11873019,  the  CAS Grand No. QYZDJ-SSW-SLH047, and the  Key Laboratory for Radio Astronomy, CAS.
 
\end{acknowledgements}

\bibliographystyle{aa} 
\bibliography{refGAIADIS} 

\longtab{
\setlength{\tabcolsep}{2.5pt}
\begin{landscape}
\begin{longtable}{ccccccccccccccp{4.5 cm}}
\caption{\label{tab:distance} The molecular cloud distances}\\
\hline\hline
  Cloud $^{\rm a}$  &  $l$$^{\rm a}$    &  $b$$^{\rm a}$   & $\rm D$$^{\rm b}$   & $\upsigma_1$$^{\rm b}$  &$\upmu_2$$^{\rm b}$& $\upsigma_2$$^{\rm b}$&  {Center$^{\rm c}$ } &  Extension$^{\rm c}$ & Noise$^{\rm d}$ &Signal$^{\rm d}$ & Cutoff&  N$^{\rm e}$    &  D$_{\rm Schlafly}$$^{\rm f}$ & Note$^{\rm g}$  \\
  &  (\deg) &  (\deg)  & (pc)   & (mag)&(mag)&(mag)& (\deg, \deg)  & (\deg, \deg)& (mJy/sr) & (mJy/sr) & (pc) &  &  (pc)\\
   (1) &  (2) &  (3)  & (4)   & (5) &(6) &(7) & (8)  & (9)& (10) & (11) & (12) &(13)   &  (14)&  (15)\\
\hline
\endfirsthead
\caption{continued.}\\
\hline\hline
  Cloud $^{\rm a}$  &  $l$$^{\rm a}$    &  $b$$^{\rm a}$   & $\rm D$$^{\rm b}$   & $\upsigma_1$$^{\rm b}$  &$\upmu_2$$^{\rm b}$& $\upsigma_2$$^{\rm b}$&  {Center$^{\rm c}$ } &  Extension$^{\rm c}$ & Noise$^{\rm d}$ &Signal$^{\rm d}$ & Cutoff&  N$^{\rm e}$    &  D$_{\rm Schlafly}$$^{\rm f}$ & Note$^{\rm g}$  \\
  &  (\deg) &  (\deg)  & (pc)   & (mag)&(mag)&(mag)& (\deg, \deg)  & (\deg, \deg)& (mJy/sr) & (mJy/sr) & (pc) &  &  (pc)\\
   (1) &  (2) &  (3)  & (4)   & (5) &(6) &(7) & (8)  & (9)& (10) & (11) & (12) &(13)   &  (14)&  (15)\\

\hline
\endhead
\hline

\endfoot
\hline
\insertTableNotes  
\endlastfoot
MBM155         &     1.6 &   -21.3 & $  148_{-3  }^{+3  }$   & $0.146_{-0.030}^{+0.028}$ & $0.378_{-0.088}^{+0.083}$ & $0.473_{-0.063}^{+0.057}$ &   (0.7, -20.8) &   (5.0,  8.7) &   3.9 &   8.3 &   300 &   626 &Y                     & MBM154,153 \\ 
MBM35          &     6.6 &    38.1 & $  107_{-4  }^{+3  }$   & $0.251_{-0.052}^{+0.047}$ & $0.502_{-0.063}^{+0.057}$ & $0.438_{-0.049}^{+0.049}$ &   (5.4,  38.1) &   (6.6,  8.2) &   4.2 &   8.2 &   300 &   617 &$89_{-25}^{+17}$      & MBM38,36,37,34,CB 63 \\ 
MBM145         &     8.5 &    21.9 & $  123_{-4  }^{+3  }$   & $0.282_{-0.050}^{+0.050}$ & $0.758_{-0.055}^{+0.058}$ & $0.503_{-0.047}^{+0.048}$ &  (11.3,  23.2) &   (9.2,  7.1) &  10.8 &  16.4 &   300 &   677 &$152_{-25}^{+19}$     & MBM146,149,148,139,150 \\ 
MBM151         &    21.5 &    20.9 & $  138_{-2  }^{+2  }$   & $0.279_{-0.036}^{+0.038}$ & $0.862_{-0.048}^{+0.047}$ & $0.472_{-0.044}^{+0.039}$ &  (20.8,  21.8) &   (6.4,  9.4) &   5.9 &  11.7 &   300 &   766 &$122_{-8}^{+8}$       &  \\ 
MBM159         &    27.4 &   -21.1 & $  133_{-3  }^{+3  }$   & $0.183_{-0.039}^{+0.035}$ & $0.330_{-0.065}^{+0.070}$ & $0.430_{-0.049}^{+0.049}$ &  (27.2, -21.3) &   (6.5,  9.4) &   5.1 &   7.3 &   300 &   786 &$143_{-10}^{+8}$      & MBM158 \\ 
Aquila S       &    38.0 &   -17.0 & $  141_{-11 }^{+13 }$   & $0.395_{-0.062}^{+0.064}$ & $0.666_{-0.046}^{+0.044}$ & $0.394_{-0.036}^{+0.038}$ &  (38.0, -17.9) &   (6.1,  7.7) &   5.0 &   8.9 &   300 &   732 &114$\pm$18            &  \\ 
MBM46          &    40.5 &   -35.5 & $  541_{-10 }^{+12 }$   & $0.335_{-0.015}^{+0.016}$ & $0.425_{-0.021}^{+0.022}$ & $0.364_{-0.018}^{+0.017}$ &  (41.7, -35.4) &   (8.6,  5.2) &   2.7 &   4.9 &  1000 &  5130 &$490_{-23}^{+25}$     & MBM47,48 \\ 
MBM160         &    44.0 &   -23.3 & $  717_{-11 }^{+11 }$   & $0.379_{-0.009}^{+0.009}$ & $0.360_{-0.008}^{+0.008}$ & $0.319_{-0.007}^{+0.007}$ &  (45.9, -21.3) &   (6.7, 10.2) &   3.4 &   5.0 &  1500 & 22411 &Y                     &  \\ 
Hercules       &    44.0 &     8.6 & $  229_{-3  }^{+4  }$   & $0.303_{-0.040}^{+0.040}$ & $1.171_{-0.052}^{+0.054}$ & $0.475_{-0.048}^{+0.045}$ &  (45.5,   9.1) &   (6.0,  2.6) &  12.4 &  16.9 &   500 &   575 &193$\pm$15            &  \\ 
Pegasus        &    92.2 &   -34.7 & $  268_{-8  }^{+6  }$   & $0.251_{-0.030}^{+0.026}$ & $0.457_{-0.056}^{+0.054}$ & $0.439_{-0.044}^{+0.041}$ &  (91.6, -32.8) &   (4.5,  6.8) &   2.5 &   5.4 &   500 &  1175 &230$\pm$24            & MBM53 \\ 
MBM54          &    93.0 &   -37.5 & $  257_{-7  }^{+8  }$   & $0.257_{-0.033}^{+0.037}$ & $0.477_{-0.068}^{+0.069}$ & $0.415_{-0.058}^{+0.056}$ &  (94.5, -37.8) &   (5.8,  3.0) &   2.7 &   5.4 &   500 &   591 &$231_{-12}^{+11}$     &  \\ 
Lacerta        &    96.0 &   -12.0 & $  512_{-6  }^{+6  }$   & $0.364_{-0.019}^{+0.021}$ & $0.757_{-0.021}^{+0.021}$ & $0.392_{-0.019}^{+0.019}$ &  (96.1, -12.3) &   (3.1,  6.5) &   5.5 &   8.9 &  1000 &  2758 &498$\pm$64            &  \\ 
MBM56          &   103.1 &   -26.1 & $  288_{-7  }^{+7  }$   & $0.334_{-0.029}^{+0.031}$ & $0.390_{-0.022}^{+0.022}$ & $0.351_{-0.019}^{+0.019}$ & (102.5, -27.4) &   (5.4,  5.2) &   2.9 &   4.5 &  1000 &  3122 &$227_{-17}^{+17}$     & G102-27 \\ 
MBM157         &   103.2 &    22.7 & $  382_{-4  }^{+5  }$   & $0.285_{-0.014}^{+0.014}$ & $0.463_{-0.012}^{+0.012}$ & $0.344_{-0.010}^{+0.010}$ & (100.3,  23.8) &   (9.5,  7.0) &   2.0 &   4.2 &  1000 &  8883 &$325_{-9}^{+11}$      & MBM156 \\ 
Cepheus(Far)   &   107.0 &    12.0 & $ 1043_{-7  }^{+6  }$   & $0.454_{-0.015}^{+0.014}$ & $2.143_{-0.018}^{+0.017}$ & $0.382_{-0.013}^{+0.014}$ & (107.5,  13.1) &   (8.7,  4.5) &  16.1 &  24.2 &  2000 &  9048 &909$\pm$108           &  \\ 
MBM162         &   111.7 &    20.1 & $  377_{-3  }^{+3  }$   & $0.399_{-0.025}^{+0.025}$ & $0.971_{-0.018}^{+0.019}$ & $0.374_{-0.017}^{+0.017}$ & (111.7,  21.3) &   (4.4,  5.1) &   4.8 &   9.1 &  1000 &  2523 &$304_{-9}^{+8}$       &  \\ 
Cepheus        &   113.0 &    17.0 & $  346_{-3  }^{+3  }$   & $0.350_{-0.024}^{+0.024}$ & $1.454_{-0.024}^{+0.024}$ & $0.487_{-0.021}^{+0.020}$ & (113.8,  17.1) &   (3.9,  6.1) &   8.4 &  14.5 &  1000 &  2347 &351$\pm$50            &  \\ 
MBM163         &   115.8 &    20.2 & $  352_{-2  }^{+2  }$   & $0.273_{-0.019}^{+0.018}$ & $0.840_{-0.013}^{+0.012}$ & $0.339_{-0.012}^{+0.012}$ & (117.0,  21.8) &   (7.2,  5.8) &   4.7 &   7.6 &  1000 &  4266 &$293_{-7}^{+6}$       & MBM161,166,165,164 \\ 
MBM2           &   117.4 &   -52.3 & $  261_{-7  }^{+8  }$   & $0.215_{-0.020}^{+0.020}$ & $0.339_{-0.023}^{+0.025}$ & $0.411_{-0.020}^{+0.016}$ & (111.5, -51.2) &  (17.6,  7.2) &   2.8 &   4.2 &  1000 &  4791 &$206_{-12}^{+14}$     &  \\ 
HSVMT 03       &   120.5 &    29.6 & $  374_{-9  }^{+9  }$   & $0.344_{-0.028}^{+0.028}$ & $0.596_{-0.017}^{+0.019}$ & $0.325_{-0.016}^{+0.018}$ & (118.9,  29.0) &   (5.7,  5.5) &   3.8 &   5.5 &  1000 &  2576 &Y                     & HSVMT 01 \\ 
HSVMT 5        &   121.1 &    21.8 & $  352_{-3  }^{+3  }$   & $0.279_{-0.019}^{+0.021}$ & $0.849_{-0.014}^{+0.013}$ & $0.317_{-0.012}^{+0.013}$ & (119.4,  21.6) &   (7.0,  4.7) &   4.6 &   7.5 &  1000 &  3549 &                      & MBM166,165,164 \\ 
HSVMT 08       &   122.5 &    29.0 & $  347_{-7  }^{+8  }$   & $0.235_{-0.044}^{+0.043}$ & $0.759_{-0.031}^{+0.030}$ & $0.295_{-0.029}^{+0.030}$ & (122.3,  29.9) &   (2.7,  3.3) &   4.8 &   7.0 &  1000 &   623 &Y                     &  \\ 
HSVMT 12       &   125.2 &    32.5 & $  361_{-7  }^{+5  }$   & $0.261_{-0.016}^{+0.015}$ & $0.550_{-0.019}^{+0.017}$ & $0.427_{-0.016}^{+0.014}$ & (127.7,  29.6) &   (6.8,  9.1) &   1.4 &   4.1 &  1000 &  5868 &Y                     & HSVMT 14 \\ 
Polaris        &   126.0 &    21.2 & $  489_{-9  }^{+10 }$   & $0.466_{-0.064}^{+0.068}$ & $0.769_{-0.017}^{+0.018}$ & $0.319_{-0.017}^{+0.015}$ & (126.6,  20.7) &   (4.7,  4.2) &   4.2 &   7.7 &  1000 &  2640 &399$\pm$37            &  \\ 
Ursa Major     &   143.0 &    38.0 & $  412_{-12 }^{+20 }$   & $0.281_{-0.021}^{+0.023}$ & $0.389_{-0.038}^{+0.035}$ & $0.411_{-0.028}^{+0.028}$ & (142.3,  36.1) &   (4.9,  7.7) &   1.3 &   3.2 &  1000 &  2411 &347$\pm$48            & MBM29,28,27,30 \\ 
MBM31          &   146.4 &    39.6 & $  400_{-8  }^{+8  }$   & $0.306_{-0.030}^{+0.032}$ & $0.534_{-0.036}^{+0.038}$ & $0.379_{-0.031}^{+0.034}$ & (147.2,  39.2) &   (4.0,  4.7) &   2.6 &   4.3 &  1000 &  1145 &$325_{-26}^{+27}$     & MBM32,HSVMT 24 \\ 
Cam            &   147.5 &    17.8 & $  225_{-6  }^{+4  }$   & $0.248_{-0.034}^{+0.034}$ & $0.429_{-0.055}^{+0.054}$ & $0.428_{-0.043}^{+0.044}$ & (146.6,  18.0) &   (6.6,  4.1) &   4.5 &   6.4 &   500 &   995 &263$\pm$88            &  \\ 
MBM7           &   150.4 &   -38.1 & $  253_{-5  }^{+6  }$   & $0.320_{-0.024}^{+0.024}$ & $0.458_{-0.047}^{+0.040}$ & $0.461_{-0.032}^{+0.034}$ & (150.9, -39.6) &  (14.0,  6.7) &   3.6 &   5.9 &   500 &  2080 &$148_{-11}^{+13}$     & MBM8,6,G154.7-39.8 \\ 
HSVMT 27       &   153.6 &    36.9 & $  397_{-8  }^{+8  }$   & $0.274_{-0.021}^{+0.020}$ & $0.353_{-0.034}^{+0.035}$ & $0.375_{-0.027}^{+0.026}$ & (152.9,  37.4) &   (6.0,  6.0) &   2.1 &   3.3 &  1000 &  2204 &Y                     &  \\ 
MBM26          &   156.4 &    32.6 & $  404_{-10 }^{+10 }$   & $0.297_{-0.026}^{+0.024}$ & $0.354_{-0.046}^{+0.043}$ & $0.421_{-0.034}^{+0.030}$ & (158.1,  33.8) &   (6.1,  4.8) &   2.1 &   3.3 &  1000 &  1802 &Y                     & HSVMT 28 \\ 
California     &   157.0 &   -12.0 & $  500_{-7  }^{+7  }$   & $0.742_{-0.030}^{+0.034}$ & $2.576_{-0.024}^{+0.024}$ & $0.209_{-0.020}^{+0.021}$ & (161.2,  -8.2) &   (8.2,  2.9) &  18.9 &  30.3 &  1000 &  1641 &414$\pm$27            &  \\ 
MBM11          &   158.0 &   -35.1 & $  279_{-3  }^{+3  }$   & $0.313_{-0.025}^{+0.025}$ & $0.861_{-0.046}^{+0.044}$ & $0.482_{-0.039}^{+0.038}$ & (159.3, -34.6) &   (7.2,  7.4) &   4.5 &   8.9 &   500 &  1206 &$185_{-20}^{+21}$     & MBM13,12,14 \\ 
UT 8d          &   158.2 &   -26.3 & $  289_{-2  }^{+3  }$   & $0.368_{-0.024}^{+0.027}$ & $1.318_{-0.023}^{+0.020}$ & $0.469_{-0.018}^{+0.018}$ & (158.1, -24.0) &   (5.5,  7.8) &   7.1 &  13.0 &  1000 &  2628 &Y                     & MBM101,103,102,104 \\ 
Perseus        &   160.0 &   -20.0 & $  310_{-4  }^{+4  }$   & $0.595_{-0.049}^{+0.046}$ & $2.059_{-0.041}^{+0.039}$ & $0.460_{-0.034}^{+0.034}$ & (159.8, -19.3) &   (4.4,  4.9) &  10.2 &  25.1 &  1000 &   980 &280$\pm$39            & MBM101,103,102,104,IC 348 \\ 
G161.9-43.3    &   161.9 &   -43.3 & $  329_{-7  }^{+7  }$   & $0.361_{-0.023}^{+0.023}$ & $0.586_{-0.017}^{+0.017}$ & $0.343_{-0.015}^{+0.016}$ & (164.0, -44.9) &  (11.2,  6.3) &   4.0 &   6.7 &  1000 &  3314 &Y                     &  \\ 
UT 4           &   165.3 &   -26.4 & $  294_{-9  }^{+13 }$   & $0.258_{-0.065}^{+0.066}$ & $0.836_{-0.032}^{+0.033}$ & $0.323_{-0.031}^{+0.030}$ & (164.0, -26.8) &   (4.0,  2.0) &   6.2 &   9.0 &  1000 &   602 &Y                     & UT 3 \\ 
MBM17          &   167.5 &   -26.6 & $  371_{-14 }^{+11 }$   & $0.524_{-0.038}^{+0.037}$ & $0.830_{-0.020}^{+0.019}$ & $0.316_{-0.017}^{+0.020}$ & (169.3, -27.5) &   (5.5,  4.0) &   7.1 &   9.7 &  1000 &  1855 &$165_{-14}^{+16}$     & UT 6,UT 5 \\ 
3C 75.0        &   170.3 &   -44.9 & $  343_{-11 }^{+8  }$   & $0.406_{-0.022}^{+0.022}$ & $0.679_{-0.019}^{+0.016}$ & $0.369_{-0.015}^{+0.016}$ & (170.6, -45.8) &   (9.0,  9.5) &   3.2 &   7.5 &  1000 &  3705 &Y                     & ir1 \\ 
MBM16          &   171.7 &   -37.7 & $  150_{-6  }^{+6  }$   & $0.180_{-0.031}^{+0.029}$ & $0.357_{-0.098}^{+0.091}$ & $0.523_{-0.059}^{+0.063}$ & (167.9, -36.9) &  (10.6,  6.6) &   5.8 &  11.2 &   300 &   817 &$147_{-9}^{+10}$      &  \\ 
Taurus         &   175.3 &   -16.2 & $  145_{-16 }^{+12 }$   & $0.441_{-0.226}^{+0.157}$ & $1.219_{-0.220}^{+0.172}$ & $0.664_{-0.119}^{+0.153}$ & (170.2, -16.2) &  (12.5,  6.7) &   9.1 &  19.1 &   300 &   828 &138$\pm$15            &  \\ 
MBM18          &   189.1 &   -36.0 & $  161_{-2  }^{+2  }$   & $0.204_{-0.024}^{+0.022}$ & $0.305_{-0.056}^{+0.062}$ & $0.412_{-0.042}^{+0.041}$ & (190.2, -37.1) &   (8.5, 11.4) &   4.8 &   9.3 &   300 &  1268 &$166_{-17}^{+18}$     & 3C 105.0 \\ 
Orion Lam      &   195.5 &   -13.7 & $  405_{-4  }^{+4  }$   & $0.350_{-0.017}^{+0.016}$ & $0.739_{-0.020}^{+0.021}$ & $0.499_{-0.018}^{+0.016}$ & (196.0, -16.6) &   (5.4,  8.7) &   5.4 &  13.2 &  1000 &  5616 &418$\pm$34            &  \\ 
Gemini         &   200.0 &    12.0 & $  725_{-18 }^{+16 }$   & $0.344_{-0.015}^{+0.017}$ & $0.510_{-0.019}^{+0.018}$ & $0.386_{-0.015}^{+0.015}$ & (198.4,  12.7) &   (5.8,  5.9) &   3.2 &   5.2 &  1500 &  6351 &Y                     &  \\ 
Mon OB1        &   201.0 &     1.0 & $  722_{-4  }^{+4  }$   & $0.336_{-0.009}^{+0.009}$ & $0.857_{-0.028}^{+0.027}$ & $0.781_{-0.020}^{+0.020}$ & (201.4,   5.0) &   (4.0, 10.6) &   4.8 &  18.0 &  1500 & 13448 &886$\pm$13            &  \\ 
NGC 2068       &   205.3 &   -14.3 & $  412_{-4  }^{+4  }$   & $0.385_{-0.040}^{+0.042}$ & $2.148_{-0.077}^{+0.082}$ & $0.486_{-0.065}^{+0.066}$ & (204.6, -14.1) &   (2.5,  2.0) &  11.2 &  27.7 &  1000 &   403 &                      &  \\ 
Rosette        &   206.8 &    -1.2 & $ 1460_{-17 }^{+19 }$   & $0.280_{-0.015}^{+0.015}$ & $2.297_{-0.038}^{+0.042}$ & $0.490_{-0.033}^{+0.035}$ & (206.4,  -2.3) &   (4.7,  3.2) &  27.4 &  56.4 &  3000 &  6017 &1483$\pm$80           &  \\ 
L1641          &   212.5 &   -19.0 & $  408_{-4  }^{+4  }$   & $0.275_{-0.030}^{+0.029}$ & $1.966_{-0.083}^{+0.079}$ & $0.724_{-0.076}^{+0.074}$ & (213.5, -19.9) &   (3.0,  3.2) &   7.5 &  30.0 &  1000 &   828 &                      &  \\ 
Mon R2         &   215.3 &   -12.9 & $  862_{-10 }^{+10 }$   & $0.327_{-0.027}^{+0.028}$ & $2.161_{-0.056}^{+0.050}$ & $0.527_{-0.044}^{+0.046}$ & (214.0, -12.3) &   (4.6,  4.4) &  13.7 &  26.2 &  1500 &  2204 &933$\pm$125           &  \\ 
ir2            &   235.0 &    37.0 & $  665_{-23 }^{+25 }$   & $0.328_{-0.017}^{+0.017}$ & $0.284_{-0.038}^{+0.042}$ & $0.405_{-0.028}^{+0.028}$ & (236.4,  38.8) &   (6.5,  5.7) &   1.7 &   3.1 &  1500 &  2877 &Y                     & 235.9+38.2 \\ 
Lupus          &   342.5 &     9.0 & $  168_{-4  }^{+4  }$   & $0.273_{-0.036}^{+0.038}$ & $1.215_{-0.043}^{+0.044}$ & $0.644_{-0.036}^{+0.037}$ & (342.3,   9.8) &   (5.0,  6.5) &  14.7 &  25.7 &   500 &  1427 &                      &  \\ 
Ophiuchus      &   355.2 &    16.0 & $  131_{-2  }^{+2  }$   & $0.201_{-0.047}^{+0.047}$ & $1.316_{-0.092}^{+0.087}$ & $0.698_{-0.074}^{+0.074}$ & (351.3,  16.5) &   (7.3,  6.2) &  14.6 &  38.6 &   300 &   443 &121$\pm$9             &  \\ 
MBM128         &   355.6 &    20.6 & $  128_{-3  }^{+2  }$   & $0.283_{-0.055}^{+0.056}$ & $0.848_{-0.121}^{+0.115}$ & $0.713_{-0.088}^{+0.092}$ & (357.0,  21.6) &   (5.4,  5.9) &  11.6 &  21.8 &   300 &   395 &$134_{-11}^{+11}$     & MBM130,133,127, 126,125,129,131 \\

\end{longtable}
\end{landscape}
 
}
\begin{TableNotes}
\footnotesize
\item [a]  Molecular clouds  (1) and their Galactic coordinates (2 and 3), most of which are cataloged by \citet{1985ApJ...295..402M},  \citet{1996ApJS..106..447M}, and \citet{2014ApJ...786...29S}.  
\item [b]  The mean and 95\% HPD errors of the distance (4),  the extinction \ag\ dispersion of foreground stars (5), the extinction \ag\ of background stars (6), and their dispersion (7).  The distance systematic error is about 5\%, which is not included in this table. 
\item [c]   The  center (8) and angular extinction (9) of region boxes in galactic coordinates, not their actual angular sizes.   Those region boxes contain both on- and off-cloud \textit{Gaia} DR2 stars.  
\item [d] The noise (10) and signal threshold (11)  used to classified on- and off-molecular clouds. 
\item [e] The number  of  on-cloud \textit{Gaia} stars (13)  nearer than the distance cutoff (12)  .
\item [f]  Distances in  \citet{2014ApJ...786...29S} (14). Because distances of multiple positions are provided for molecular clouds in Table 1 of \citet{2014ApJ...786...29S}, we averaged the distances of those molecular clouds  weighted by  their errors. Molecular clouds that not measured by \citet{2014ApJ...786...29S} or other reliable methods  are marked with Y, 15 in total.
\item [g] Other molecular clouds covered by the region box (15).
\item [h] The far component towards the Cepheus molecular clouds.
\end{TableNotes}

\end{document}